\documentclass[twocolumn,aps,prd,reprint,superscriptaddress,nofootinbib,letterpaper]{revtex4}

\usepackage{amsmath}
\usepackage{amsfonts}
\usepackage{amssymb}
\usepackage{graphicx, nicefrac}
\usepackage{epsfig}
\usepackage{color}
\usepackage{multirow}
\usepackage{todonotes}
\usepackage{hyperref}
\usepackage[normalem]{ulem}

\def\bea{\begin{eqnarray}}
\def\eea{\end{eqnarray}}
\def\bef{\begin{flalign}}
\def\eef{\end{flalign}}
\def\nn{\nonumber}

\def\vk{\vec{k}}
\def\vx{\vec{x}}

\def\({\left(}
\def\){\right)}
\def\[{\left[}
\def\]{\right]}
\def\<{\left\langle}
\def\>{\right\rangle}

\begin{document}


\title{Linear growth of structure in massive gravity}

\author{Ekapob Kulchoakrungsun}
\email{ek2897@nyu.edu}
\affiliation{Center for Cosmology and Particle Physics, Department of Physics, New York University, 726 Broadway, New York, NY 10003, USA}

\author{Ananya Mukherjee}
\email{ananya\_mukherjee@student.uml.edu}
\affiliation{Department of Physics and Applied Physics, University of Massachusetts, Lowell, MA 01854, USA}

\author{Nishant Agarwal}
\email{nishant\_agarwal@uml.edu}
\affiliation{Department of Physics and Applied Physics, University of Massachusetts, Lowell, MA 01854, USA}

\author{Anthony R. Pullen}
\email{anthony.pullen@nyu.edu}
\thanks{\\ {\bf E. K. and A. M. contributed equally to this work.}}
\affiliation{Center for Cosmology and Particle Physics, Department of Physics, New York University, 726 Broadway, New York, NY 10003, USA}
\affiliation{Center for Computational Astrophysics, Flatiron Institute, New York, NY 10010, USA}

\date{\today} 

\begin{abstract}
We study background dynamics and the growth of matter perturbations in the extended quasidilaton setup of massive gravity. For the analysis of perturbations, we first choose a scalar field matter component and obtain the conditions under which all scalar perturbations are stable. We work in unitary gauge for the matter field, which allows us to directly map to known results in the limit of general relativity. By performing a parameter search, we find that the perturbations are unstable in general, while a particular choice of potential, where the scalar field effectively behaves like pressureless matter, allows for stable perturbations. We next consider the growth of matter perturbations in a cold dark matter-dominated Universe. Working in conformal Newtonian gauge, we obtain evolution equations for various observables including the growth factor and growth rate, and find scale-independent growth in the quasistatic and subhorizon approximations. We finally show how the Hubble parameter and matter perturbations evolve in massive gravity for a specific choice of parameter values, and how this evolution compares to the standard cosmological model consisting of a cosmological constant and cold dark matter.
\end{abstract}

\maketitle


\section{Introduction}

The origin of the current accelerated expansion of the Universe is a mystery of modern cosmology. While one possibility is to have an additional source of energy, commonly called dark energy and well-fit by a cosmological constant, another equally compelling possibility is that the theory of gravity differs from general relativity (GR) on large scales. Along the lines of the latter, an interesting question is whether the graviton can be massive, with a mass of the order of the Hubble constant today. Early attempts to make the graviton massive were restricted to a linear theory of massive gravity \cite{Fierz:1939ix}, and a nonlinear theory, commonly called the de Rham-Gabadadze-Tolley (dRGT) theory, was successfully constructed relatively recently in \cite{deRham:2010ik,deRham:2010kj}; see \cite{deRham:2014zqa,Hinterbichler:2011tt} for detailed reviews.

The dRGT theory is free of the Boulware-Deser ghost \cite{Boulware:1972yco}, but does not admit a viable Friedmann-Lemaître-Robertson-Walker (FLRW) cosmology \cite{DAmico:2011eto}. One potential resolution of this problem is to add to the theory a quasidilaton field that realizes a new global symmetry \cite{DAmico:2012hia}, the cosmological implications of which were studied in \cite{Gannouji:2013rwa}. However, this modification was also found to have unstable perturbations around a self-accelerating background \cite{Gumrukcuoglu:2013nza}. A further extension, where one also allows for a new type of coupling between the massive graviton and the quasidilaton \cite{DeFelice:2013tsa}, on the other hand, does allow for a stable self-accelerated solution.

Background dynamics in the extended quasidilaton setup with a kinetic term for the quasidilaton field were studied in \cite{DeFelice:2013tsa,Kahniashvili:2014wua} and the evolution of perturbations has been studied in, for example, \cite{Motohashi:2014una,Heisenberg:2015voa,Gumrukcuoglu:2016hic}. It turns out that the Boulware-Deser ghost is not guaranteed to be absent away from the self-accelerating attractor in this case \cite{Kluson:2013jea,Mukohyama:2013raa,Anselmi:2017hwr,Golovnev:2017zxk}, with one possible resolution being to remove the quasidilaton kinetic term. The resulting model without the kinetic term was shown to pass all perturbative stability tests with a self-accelerating background solution in the absence of matter in \cite{Gumrukcuoglu:2017ioy}. In this paper, we are interested in understanding how scalar perturbations evolve in this model of massive gravity in the presence of matter.

We first find the conditions for stable scalar perturbations in the presence of a scalar field matter component. We perform this calculation in unitary gauge for the matter field, which we find to be the most convenient choice and also allows us to directly map our result to the action for inflationary perturbations in the limit of GR. We next find the equations of motion for all propagating scalar perturbations in the presence of cold dark matter (CDM). We perform this calculation in the commonly-used conformal Newtonian gauge and finally obtain equations for various observables including the growth factor and growth rate. Growth of structure has been explored in different models of massive gravity in earlier works as well, for example, bi-metric massive gravity \cite{Solomon:2014dua}, generalized massive gravity \cite{Kenna-Allison:2020egn}, projected massive gravity \cite{Manita:2022}, and the minimal theory of massive gravity \cite{DeFelice:2021trp}. We find scale-independent growth in the quasistatic and subhorizon approximations, in agreement with the latter two references that also had a single dynamical metric. While the complexity of this theory limits us from going beyond the study of linear perturbations, this scale-independent growth is an important difference from $f(R)$ theories of gravity, where one finds a scale-dependent growth of structure \cite{Zhang:2005vt,Bean:2006up}.

We lastly solve the evolution equations for the background and perturbations numerically. We find that the stability conditions are not satisfied in general, but can be satisfied for a particular choice of potential where the scalar field effectively behaves like pressureless matter. Choosing parameter values that provide a reasonable fit (not the best fit, which would require a more detailed analysis) to background and growth of structure data, we show how the Hubble parameter and matter perturbations evolve in massive gravity compared to the standard $\Lambda$CDM model, $\Lambda$ being a cosmological constant. For the parameter values that we consider, we find good agreement with the background evolution in $\Lambda$CDM but differences in the growth of matter perturbations at late times.

The paper is organized as follows. We start with an overview of the extended quasidilaton setup of massive gravity in Sec.\ \ref{section:Extended_quasidilaton_theory}. In Sec.\ \ref{section:Background_evolution}, we obtain evolution equations for the background in the presence of matter and show that the evolution is restricted to one of two branches, identical to what one finds in the absence of matter. We introduce scalar perturbations in Sec.\ \ref{section:Cosmological_perturbations}, and check for the stability of propagating perturbations in each branch in Sec.\ \ref{section:Kinetic_matrix}. In Sec.\ \ref{section:Growth_rate}, we solve the equations of motion for the perturbations to find how CDM perturbations grow in each branch. We present numerical solutions to the equations of motion and compare the resulting evolution of the Hubble parameter and matter perturbations to $\Lambda$CDM in Sec.\ \ref{section:Figures}. We end with a summary and discussion of our results in Sec.\ \ref{section:Discussion}. The first appendix shows that the two gauge choices made in the paper are valid and the second appendix contains expressions for certain quantities that appear in Sec.\ \ref{section:Growth_rate}.

A note on our notation: Greek indices indicate time and space coordinates and take the values $0$-$3$, Latin indices indicate space coordinates and take the values $1$-$3$, and our metric signature is mostly plus.



\section{Extended quasidilaton theory}
\label{section:Extended_quasidilaton_theory}

The dRGT theory introduces the graviton mass in a covariant way by means of the St\"{u}ckelberg mechanism. The four St\"{u}ckelberg fields $\phi^{\alpha}$ together generate the nondynamical metric
\bea
    f_{\mu\nu} \, = \, \eta_{\alpha\beta} \partial_{\mu} \phi^{\alpha} \partial_{\nu} \phi^{\beta} \, ,
\eea
where $\eta_{\alpha\beta}$ is the Minkowski metric and the derivative is with respect to $x^{\mu} = (t,\vx)$. The tensor $\( \sqrt{g^{-1}f} \)^{\mu}_{\ \nu}$ forms the basic building block of the mass term, $g_{\alpha\beta}$ being the dynamical spacetime metric. Since the dRGT theory does not admit a viable FLRW cosmology \cite{DAmico:2011eto}, as mentioned in the Introduction, we focus here on its quasidilaton extension, where a quasidilaton field $\sigma$ \cite{DAmico:2012hia} is coupled to the St\"{u}ckelberg fields through an extended fiducial metric \cite{DeFelice:2013tsa},
\bea
    \tilde{f}_{\mu\nu} \, = \, \eta_{\alpha\beta} \partial_{\mu} \phi^{\alpha} \partial_{\nu} \phi^{\beta} - \frac{\alpha_\sigma}{m^2} \partial_{\mu} (e^{-\sigma}) \partial_{\nu} (e^{-\sigma}) \, ,
\label{eq:ftilde}
\eea
$m$ being the mass of the graviton and $\sigma$ the dimensionless field $\sigma/M_{\text{Pl}}$. This transforms as $\tilde{f}_{\mu\nu} \rightarrow e^{-2\sigma_0} \tilde{f}_{\mu\nu}$ under the global transformations $\sigma \rightarrow \sigma + \sigma_0$ and $\phi^{\alpha} \rightarrow e^{-\sigma_0} \phi^{\alpha}$, with $\sigma_0$ an arbitrary constant. Following \cite{Gumrukcuoglu:2017ioy}, we also remove the canonical kinetic term for the quasidilaton field.

In this paper, we are interested in the growth of matter perturbations in the setup described above. The final action that we consider is thus
\bea
    S & = & \frac{M_\mathrm{Pl}^2}{2} \int d^4x \sqrt{-g} \[ R + 2m^2 \( \mathcal{L}_2 + \alpha_3 \mathcal{L}_3 + \alpha_4 \mathcal{L}_4 \) \] \nn \\
    & & \quad + \, \int d^4 x \sqrt{-g} \, \mathcal{L}_\mathrm{matter} \, ,
\label{eq:action}
\eea
where $R$ is the 4D Ricci scalar. The mass term is generated by the Lagrangian densities
\bea
    \mathcal{L}_2 & = & \frac{1}{2!} \( [\mathcal{K}]^2 - [\mathcal{K}^2] \) , \\
    \mathcal{L}_3 & = & \frac{1}{3!} \( [\mathcal{K}]^3 - 3[\mathcal{K}][\mathcal{K}^2] + 2 [\mathcal{K}^3] \) , \\
    \mathcal{L}_4 & = & \frac{1}{4!} \big( [\mathcal{K}]^4 - 6[\mathcal{K}]^2[\mathcal{K}^2] + 3 [\mathcal{K}^2]^2 + 8[\mathcal{K}][\mathcal{K}^3] \nn \\
    & & \quad - \ 6[\mathcal{K}^4] \big) \, ,
\eea
where square brackets denote the trace, and
\bea
    \mathcal{K}^\mu_{\ \nu} \, = \, \delta^\mu_\nu - e^{\sigma} \( \sqrt{g^{-1} \tilde{f}} \)^{\mu}_{\ \nu} \, .
\eea
The matter Lagrangian density $\mathcal{L}_\mathrm{matter}$, on the other hand, gives the energy-momentum tensor in the standard way,
\bea
    T_{\mu\nu} & = & -\frac{2}{\sqrt{-g}} \frac{\delta}{\delta g^{\mu\nu}} \( \sqrt{-g} \, \mathcal{L}_\mathrm{matter} \) .
\label{eq:stress-energy tensor}
\eea
We can now use the action in Eq.\ (\ref{eq:action}) to find both the background evolution, which we review in the next section, and the evolution of perturbations, which we study in the remainder of the paper.



\section{Background evolution}
\label{section:Background_evolution}

Let us first consider the Einstein-Hilbert part of the action. For this, we choose to work in the Arnowitt-Deser-Misner (ADM) formalism, since this allows us to easily identify the boundary term, as explained further below. In this formalism, spacetime is foliated into spacelike hypersurfaces and is described in terms of a spatial metric $h_{ij}$, lapse function $N$, and shift vector $N^i$, with the spacetime interval written as
\bea
    ds^2 & = & -N^2 dt^2 + h_{ij} \(N^i dt + dx^i \) \( N^j dt + dx^j \) .
\eea
The functions here are related to components of the original metric $g_{\mu\nu}$ as
\bea
    g_{\mu\nu} \, = \, \( {\begin{array}{cc}
        -N^2 + N^k N_k & N_j \\
        N_i & h_{ij} \\
    \end{array}} \) ,
\eea
while for the inverse metric, we have
\bea
    g^{\mu\nu} \, = \, \frac{1}{N^2} \( {\begin{array}{cc}
        1 & N^j \\
        N^i & N^2 h^{ij} - N^i N^j \\
    \end{array}} \) .
\eea
We can now use the Gauss-Codazzi equation to relate the 4D Ricci scalar to the 3D Ricci scalar $R^{(3)}$,
\bea
    R \, = \, R^{(3)} + K^i_j K^j_i - [K]^2 + 2\nabla_{\mu} \( [K] n^{\mu} - n^{\nu} \nabla_{\nu} n^{\mu} \) , \nn \\
\label{eq:gc}
\eea
where $\nabla_{\mu}$ is the covariant derivative with respect to the metric $g_{\mu\nu}$, $K_{ij} = \frac{1}{2N} \big( \dot{h}_{ij} - D_i N_j - D_j N_i \big)$ is the extrinsic curvature, with the dot denoting a derivative with respect to $t$ and $D_i$ being the covariant derivative with respect to the induced metric $h_{ij}$, and $n^{\mu} = \frac{1}{N} \( 1, -N^i \)$ is the normal to the spatial slice. The boundary term is now easily identified as the last term in Eq.\ (\ref{eq:gc}) and is removed by adding an appropriate Gibbons-Hawking-York boundary term. We find this identification convenient to unambiguously construct the second-order action in the perturbations later in the paper. As for the background, we choose an FLRW cosmology with $N_i$ set to zero and $h_{ij} = a^2(t) \delta_{ij}$, where $a(t)$ is the scale factor.

Let us next consider the massive gravity part of the action. Consistent with a homogeneous and isotropic background cosmology, we choose the background St\"{u}ckelberg fields to be 
\bea
    \phi^{\alpha} & = & \delta^{\alpha}_0 \phi(t) + \delta^{\alpha}_i x^i \, ,
\eea
where $\phi(t)$ is some function of $t$. Note that we will not work in unitary gauge for the St\"{u}ckelberg fields and will thus allow their perturbations to be functions of spacetime later in the paper. The background quasidilaton field is similarly chosen to be a function of $t$ only. Using this in Eq.\ (\ref{eq:ftilde}) gives the following fiducial spacetime interval,
\bea
    ds^2_{\tilde{f}} & = & -r^2(N^2/a^2) dt^2 + \delta_{ij} dx^i dx^j \, ,
\eea
where
\bea
    r^2 \frac{N^2}{a^2} & = & \dot{\phi}^2 + \frac{\alpha_\sigma}{m^2} e^{-2\sigma} \dot{\sigma}^2
\label{eq:r definition}
\eea
is the effective lapse function.

We can now obtain the zeroth order action from Eq.\ (\ref{eq:action}). We define $X = e^{\sigma}/a$ and the following combinations of background quantities as in \cite{Gumrukcuoglu:2017ioy},
\bea
    J & = & (3-2X) + (X-3)(X-1) \alpha_3 + (X-1)^2 \alpha_4 \, , \nn \\
\label{eq:def J} \\
    Q & = & (X-1) [3 - 3(X-1) \alpha_3 + (X-1)^2 \alpha_4] \, ,
\label{eq:def Q}
\eea
in terms of which the zeroth order action is given by
\bea
    S^{(0)} & = & M_{\rm Pl}^2 \int dt d^3x a^3 N \[ -3H^2 + m^2 (rQX - \rho_X) \] \nn \\ 
    & & \quad + \ S_{\rm matter} \, ,
\eea
where $H = \dot{a}/(aN)$ is the Hubble parameter and
\bea
    \rho_X & = & \frac{1}{X} \[ Q + J (X-1)^2 - X (X-1)^2 \]
\eea
can be interpreted as the contribution to the energy density from massive gravitons. The background equations of motion are obtained by varying $S^{(0)}$ with respect to the background fields $\{N(t), a(t), \phi(t), \sigma(t)\}$,
\bea
    \delta S^{(0)} & = & \int d^4x \sum_i \( \frac{\delta S^{(0)}}{\delta\Phi^i} \delta\Phi^i \) \, ,
\label{eq:varying action}
\eea
and setting the variation $\delta S^{(0)}$ to zero. (The summation here is over the four fields mentioned above and not a spatial index.) After the variation, we can either set $N$ to unity, so that the time coordinate corresponds to physical time $t$, or to $a(\tau)$, so that the time coordinate corresponds to conformal time $\tau$. We will work in conformal time below.

Assuming that matter is a perfect fluid so that its energy-momentum tensor is given by $\bar{T}^{\mu}_{\nu} = {\rm diag} ( -\bar{\rho}, \bar{p}, \bar{p}, \bar{p} )$, where a bar denotes background quantities, the resulting four independent equations describing the background evolution are 
%
\bea
     \frac{\mathcal{H}^2}{a^2} & = & \frac{m^2 \rho _X}{3} + \frac{\Bar{\rho}}{ 3M_{\text{Pl}}^2} \, ,
\label{eq:friedmann1}
\eea
%
\bea
    2\( \frac{\mathcal{H}' - \mathcal{H}^2}{a^2} \) & = & m^2 JX(r-1) - \frac{\bar{\rho} + \bar{p}}{M_{\mathrm{Pl}}^2} \, ,
\label{eq:friedmann2}
\eea
%
\bea
    \frac{d}{d\tau} \( \frac{a^4 QX\phi'}{r} \) & = & 0 \, ,
\label{eq:Stuckelberg background}
\eea
%
\bea
    \frac{\alpha_\sigma}{X a^5} \frac{d}{d\tau} \( \frac{a^3 Q \sigma'}{r} \) & = & m^2 X [3J(r-1) + 4rQ] \, , \quad
\label{eq: sigma background}
\eea
where ${\cal H} = aH$ is the conformal Hubble parameter and a prime denotes a derivative with respect to $\tau$. Combining the first and second Friedmann equations\ (\ref{eq:friedmann1}) and (\ref{eq:friedmann2}) and using the relationships $\bar{\rho}' + 3{\cal H}(\bar{\rho} + \bar{p}) = 0$, $\rho_X' = 3JX'$, and $X' = (\sigma' - {\cal H}) X$ yields the constraint
\bea
    m^2 J X \left( \sigma' - \mathcal{H}r \right) & = & 0 \, .
\label{eq:two branches eq}
\eea
This restricts the background evolution to two branches: $\sigma' = \mathcal{H}r$ (branch $1$) and $J = 0$ (branch $2$); these are identical to the two branches found in \cite{Gumrukcuoglu:2017ioy} in the absence of matter. The first branch condition leads to the evolution equation $X' = (r-1) {\cal H}X$ for $X$. In the second branch, on the other hand, $X$ is constant, so that $X' = 0$ and $\sigma' = {\cal H}$.


\section{Scalar Perturbations}
\label{section:Cosmological_perturbations}

We next consider how scalar perturbations evolve in massive gravity. We have a total of eight perturbations before making use of the gauge freedom in the theory. Four of the perturbations come from the dynamical metric,
\bea
    & & ds^2 \, = \, a^2(\tau) \bigg[ -(1+2\Phi) d\tau^2 + 2\partial_i B dx^i d\tau \nn \\
    & & \ \ + \, \bigg\{ (1 - 2\Psi) \delta_{ij} + 2 \( \partial_i \partial_j - \frac{\delta_{ij}}{3} \partial^2 \) E \bigg\} dx^i dx^j \bigg] , \quad \ \ 
\label{Eq:metric_perturbation}
\eea
where $\Phi$, $B$, $\Psi$, and $E$ are all functions of $(\tau,\vx)$. Another two come from the St\"{u}ckelberg fields,
\bea
    \delta\phi^0 & = & \Pi^0 \, , \\
    \delta\phi^i & = & \Pi^i + \partial^i \Pi_L \, .
\eea
where $\Pi^0$ and $\Pi_L$ are scalars and also functions of $(\tau,\vx)$. $\Pi^i$ is the vector part of the perturbation, that we will ignore in this paper. The next scalar perturbation is in the quasidilaton field, $\delta\sigma(\tau,\vx)$. And the last one is in the matter component, that we consider in the next two sections. We show how the perturbations transform under a coordinate transformation in appendix\ \ref{app:gauge} and identify two gauge choices that are convenient for our calculations in the following sections.


\section{Kinetic matrix}
\label{section:Kinetic_matrix}

In this section, we obtain the conditions under which the scalar perturbations that we introduce are stable. For this, we check the signs of the kinetic terms of all propagating scalars and demand that the theory does not propagate any ghosts. We choose a scalar field matter component for simplicity, denoted $\chi(\tau,\vx)$, with a Lagrangian density of the form
\bea
    \mathcal{L_\mathrm{matter}} & = & -\frac{1}{2} \left( \partial_\mu\chi \right)^2 - V(\chi) \, ,
\eea
where $(\partial_{\mu} \chi)^2 = g^{\mu\nu} (\partial_{\mu}\chi) (\partial_{\nu}\chi)$ and $V$ is some potential. We perturb the scalar field as $\chi(\tau,\vx) = \bar{\chi}(\tau) + \delta\chi(\tau,\vx)$, $\bar{\chi}(\tau)$ being the background field. The Friedmann equations describing the background evolution of the Universe are then given by Eqs.\ (\ref{eq:friedmann1}) and (\ref{eq:friedmann2}) with $\bar{\rho} = \bar{\chi}'^2/(2 a^2) + V$ and $\bar{p} = \bar{\chi}'^2/(2 a^2) - V$. We use the resulting equations to substitute for the background $V$ and $a''$ in the remainder of this section.

Let us next consider the choice of gauge for the perturbations. We find it simplest to work in unitary gauge for the matter field for the analysis in this section, and so make the gauge choice $E = 0$ and $\delta\chi = 0$; we show that this is a valid gauge choice in appendix\ \ref{app:gauge}. Note that this is also the typical gauge choice for calculating the spectrum of perturbations in single-field inflation. We thus expect our Lagrangian density to reduce to that for the primordial curvature perturbation $\zeta(\tau,\vx)$ in the limit of ``turning off" massive gravity, and show that this is indeed true later in the section.

With the gauge choice that we have made, we are left with six scalar perturbations: $\Phi$, $B$, $\Psi$, $\Pi^0$, $\Pi_L$, and $\delta \sigma$. We introduce these perturbations at the level of the action in Eq.\ (\ref{eq:action}) and write the resulting second-order action in Fourier space, using the convention that $f(\tau,\vx) = \int \frac{d^3k}{(2\pi)^3} e^{i \vk \cdot \vx} f(\tau,\vk)$. As is usual with the lapse and shift, the perturbations $\Phi$ and $B$ turn out not to have any time derivatives and lead to constraint equations. They can thus be written in terms of other perturbations and we find the solutions
\begin{widetext}
\bea 
    \Phi &=& -\frac{a M_{\text{Pl}}^2}{X \left\{2 a^2 \mathcal{H}^2 M_{\text{Pl}}^2 \left[3 a^2 J m^2 X+2 k^2 (r+1)\right]-a^4 J m^2 X \left(\Bar{\chi}'\right)^2\right\}} \left\{a^5 J^2 m^4 X^3 \left[3 (\delta \sigma +\Psi )-k^2 \Pi _L\right] \right.\nn\\
    && \left.+ \, 2 a^3 J m^2 X^2 \mathcal{H} \left(-k^2 \Pi _L'+k^2 \Pi _0 \phi '+3 \Psi '\right)+2 a^3 J k^2 m^2 X^2 \Psi +2 a k^2 \mathcal{H} \left[\delta \sigma  J \alpha _{\sigma } \sigma '+2 (r+1) X \Psi '\right]\right\} ,
\label{eq:Phi constraint} \\
    B &=& \frac{1}{X \left\{2 \mathcal{H}^2 M_{\text{Pl}}^2 \left[3 a^2 J m^2 X+2 k^2 (r+1)\right]-a^2 J m^2 X \left(\Bar{\chi}'\right)^2\right\}} \Big\{2 \mathcal{H} M_{\text{Pl}}^2 \left[-a^2 J k^2 m^2 (r+1) X^2 \Pi _L \right.\nn\\
    && \left.+ \, X \left((r+1) \left(3 a^2 J m^2 X (\delta \sigma +\Psi )+2 k^2 \Psi \right)+3 a^2 J m^2 X \mathcal{H} \Pi _L'-3 a^2 J m^2 \Pi _0 X \mathcal{H} \phi '\right)-3 \delta \sigma  J \mathcal{H} \alpha _{\sigma } \sigma '\right] \nn\\
    && +\left(\Bar{\chi}'\right)^2 \left[X \left(-a^2 J m^2 X \Pi _L'+a^2 J m^2 \Pi _0 X \phi '+2 (r+1) \Psi '\right)+\delta \sigma  J \alpha _{\sigma } \sigma '\right]\Big\} \, ,
\label{eq:B constraint}
\eea 
\end{widetext}
which reduce to the corresponding equations of \cite{Gumrukcuoglu:2017ioy} in the absence of matter on using the gauge freedom to set $\delta\sigma$ to zero instead. 

Substituting the solutions for $\Phi$ and $B$ back into the second-order action, thus integrating them out, yields an action of the form
\bea 
    S^{(2)} & = & \frac{M_{\text{Pl}}^2}{2}\int d\tau d^3k a^2 \left(Y'^T K Y' + Y'^T M Y \right.\nonumber\\
    && \quad \left.- \ Y^T M Y' - Y^T \Omega^2 Y \right),
\label{eq:kineticaction}
\eea
where $Y$ denotes the column vector $\left(\Psi, \delta\sigma, m\Pi_0, m^2\Pi_L\right)$, $K$ is the kinetic matrix, $M$ is the mixing matrix, and $\Omega^2$ is the frequency matrix. Perturbations to the St\"{u}ckelberg fields, $\Pi_0$ and $\Pi_L$, are dimensionful, and we thus multiply them with factors of $m$ to convert to dimensionless variables. This also yields a dimensionless kinetic matrix. In doing so, we are implicitly assuming the following scaling for the perturbations: $\Psi \sim \delta\sigma \sim m\Pi_0 \sim m^2 \Pi_\mathrm{L}$, which will be important when we consider the subhorizon approximation later in the paper. We will motivate this scaling when we look closely at the equations of motion in Sec.\ \ref{section:Growth_rate}.

The $4\times4$ matrix $K$ in Eq.\ (\ref{eq:kineticaction}) is actually a block diagonal matrix of the form
\bea
    K & = & \left(
    \begin{array}{cc}
    K_{2\times2}(\Psi,\Pi_L) & 0 \\
    0 & K_{2\times2}(\delta\sigma,\Pi_0) \\
\end{array}
\right) ,
\eea
where $K_{2\times2}(\Psi,\Pi_L)$, and $K_{2\times2}(\delta\sigma,\Pi_0)$ are the $2\times2$ kinetic matrices of $\( \Psi, \Pi_L \)$ and $\( \delta\sigma, \Pi_0 \)$ respectively. Components of the first block are given by
\begin{widetext}
\bea 
    K_{\Psi\Psi} &=& -\frac{2 \left(\Bar{\chi}'\right)^2 \left(3 a^2 J m^2 X+2 k^2 (r+1)\right)}{a^2 J m^2 X \left(\Bar{\chi}'\right)^2-2 \mathcal{H}^2 M_{\text{Pl}}^2 \left(3 a^2 J m^2 X+2 k^2 (r+1)\right)} \,,\\
    K_{\Psi\Pi_L} &=& \frac{2 a^2 J k^2 X \left(\Bar{\chi}'\right)^2}{a^2 J m^2 X \left(\Bar{\chi}'\right)^2-2 \mathcal{H}^2 M_{\text{Pl}}^2 \left(3 a^2 J m^2 X+2 k^2 (r+1)\right)} \,,\\
    K_{\Pi_L\Pi_L} &=& \frac{4 a^2 J k^4 X \mathcal{H}^2 M_{\text{Pl}}^2}{2 m^2 \mathcal{H}^2 M_{\text{Pl}}^2 \left(3 a^2 J m^2 X+2 k^2 (r+1)\right)-a^2 J m^4 X\left(\Bar{\chi}'\right)^2} \,,
\eea 
\end{widetext}
while those of the second block are given by
\bea 
    K_{\delta\sigma\delta\sigma} &=& \frac{Q \alpha _{\sigma } \left(a^2 m^2 r^2 X^2-\alpha _{\sigma } \left(\sigma '\right)^2\right)}{a^2 m^2 r^3 X^3} \, ,\\
    K_{\delta\sigma\Pi_0} &=& -\frac{Q \alpha _{\sigma } \sigma ' \phi '}{m r^3 X}\,,\\
    K_{\Pi_0\Pi_0} &=& \frac{a^2 Q X \left(r^2-\left(\phi '\right)^2\right)}{r^3} \, .
\eea 
In the absence of matter, $K_{\Psi\Psi}$ and all cross-terms of $\Psi$ vanish. Therefore, $\Psi$ is identified as the Boulware-Deser degree, in agreement with \cite{Gumrukcuoglu:2017ioy}.

As another check, we can obtain the limit of GR with a scalar field matter component by first setting $\alpha_\sigma \rightarrow 0$ and then $m \rightarrow 0$. Only $\Psi$ remains on doing so and, after integrating a $\Psi'\Psi = \frac{1}{2}\frac{d}{d\tau}\Psi^2$ term by parts, the second-order action becomes
\bea 
    S^{(2)} & = & \frac{1}{2} \int d\tau d^3k \left( \frac{a^4\Bar{\chi}'^2}{a'^2} \right) \left( \Psi'^2 - k^2\Psi^2 \right) .
\eea
This is in exact agreement with what one finds for the comoving curvature perturbation $\zeta$ on perturbing the action for the inflaton field. Our second-order action thus has the correct GR limit in the presence of matter.

We do not show explicit expressions for the remaining matrices in Eq.\ (\ref{eq:kineticaction}) as they are quite big, but they can be found in our {\tt Mathematica} supplement. In the following two subsections, we analyze the kinetic matrix in the two branches of solutions that the background equation (\ref{eq:two branches eq}) had yielded.

\subsection{Branch $1$: $\sigma' = \mathcal{H}r$}

Since the $4\times4$ matrix $K$ is a block diagonal matrix, the eigenvectors and eigenvalues of $K$ are simply those of each block combined. With the branch condition, we find that the $2\times2$ submatrix $K_{2\times2}(\delta\sigma,\Pi_0)$ has an extra vanishing eigenvalue, implying that a linear combination of $\delta\sigma$ and $\Pi_0$ is non-dynamical. Since we expect to find a Boulware-Deger degree, this non-dynamical field is indeed what we are looking for. To find this non-dynamical perturbation, we diagonalize the $2\times2$ submatrix $K_{2\times2}(\delta\sigma,\Pi_0)$ and use its eigenvectors to define two new perturbations $(\widetilde{\delta\sigma},\widetilde{\Pi}_0)$ as
\bea 
    \delta \sigma & = & \frac{r \mathcal{H}\widetilde{\delta\sigma} - m^2  \phi' \widetilde{\Pi} _0}{\sqrt{m^2 \left(\phi '\right)^2 + r^2 \mathcal{H}^2}} \, , \\
    \Pi_0 & = & \frac{\phi' \widetilde{\delta\sigma} + r \mathcal{H}\widetilde{\Pi}_0}{\sqrt{m^2 \left(\phi '\right)^2 + r^2 \mathcal{H}^2}} \, .
\eea
With these (time-dependent) field redefinitions, the new perturbation $\widetilde{\delta\sigma}$ has no kinetic term and is therefore non-dynamical.

The equation of motion for $\widetilde{\delta\sigma}$ is a constraint equation of the form
\bea
    \widetilde{\delta\sigma} \, = \, c_1 \Psi + c_2 \widetilde{\Pi}_0 + c_3 \Pi_L + c_4 \Psi' + c_5 \widetilde{\Pi}_0' + c_6 \Pi_L' \, ,
\label{eq:delta sigma constraint}
\eea 
where the coefficients $c_i$ are functions of background quantities. We substitute this solution back into the second-order action, whose form was shown in Eq.\ (\ref{eq:kineticaction}), after integrating a $\widetilde{\delta\sigma}' \widetilde{\delta\sigma} = \frac{1}{2} \frac{d}{d\tau} \big( \widetilde{\delta\sigma} \big)^2$ term by parts and after integrating by parts cross-terms of the form $\widetilde{\delta\sigma}' \Psi$, $\widetilde{\delta\sigma}' \widetilde{\Pi}_0$, and $\widetilde{\delta\sigma}' \Pi_L$. This results in a $3 \times 3$ kinetic matrix for the perturbations $\( \Psi, m\widetilde{\Pi}_0, m^2 \Pi_L \)$. We again do not show the resulting matrix explicitly here given its complexity, but it can be found in our {\tt Mathematica} supplement.

In the subhorizon approximation ($k/\mathcal{H} \gg 1$), it suffices to check that the final kinetic matrix has positive eigenvalues for the perturbations to be stable. It turns out to be easier to check equivalently that the kinetic matrix is positive definite, which is what we demand here. This implies that the determinants of all $n \times n$ upper-left sub-matrices must be positive. In the subhorizon approximation, this yields the following three conditions, 
\begin{widetext}
\bea 
    0 &<& \frac{\left(\Bar{\chi}'\right)^2 }{2 a^2 m^6 r X^4 \mathcal{H}^2 M_{\text{Pl}}^2} \left[a^6 \left(-m^6\right) X^5 \left(J Q \alpha _{\sigma } (-3 J r-4 Q (r-1))+2 r X (3 J+4 Q)^2\right)\right. \nn\\
    && -a^6 m^4 X^5 \mathcal{H}^2 \left(J Q \alpha _{\sigma } (-3 J r-4 Q (r-1))+2 r X (3 J+4 Q)^2\right) \nn\\
    && +a^4 m^4 X^3 \mathcal{H}^2 \alpha _{\sigma } \left(J Q \alpha _{\sigma } (J (6-9 r)-11 Q (r-1))+2 X (3 J+4 Q) (6 J (2 r-1)+14 Q r+Q)\right) \nn\\
    && +a^4 m^2 X^3 \mathcal{H}^4 \alpha _{\sigma } \left(J Q^2 (r-1) \alpha _{\sigma }+2 X (3 J+4 Q) (3 J r+2 Q r+Q)\right) \nn\\
    && -a^2 m^2 X \mathcal{H}^4 \alpha _{\sigma }^2 \left(2 X \left(9 J^2 (3 r-2)+9 J Q (7 r-2)+Q^2 (37 r+11)\right)-J Q^2 (r-1) \alpha _{\sigma }\right) \nn\\
    && \left.+2 a^2 Q X^2 \mathcal{H}^6 \alpha _{\sigma }^2 (3 J (r-1)+3 Q r+Q)+2 Q \mathcal{H}^6 \alpha _{\sigma }^3 (3 J (r-1)+3 Q r+Q)\right] \nn\\
    && - \frac{Q \alpha _{\sigma } \left(\Bar{\chi}'\right){}^4 }{2 a^2 m^6 r X^4 \mathcal{H}^2 M_{\text{Pl}}^4}\left[a^4 m^2 X^4 (3 J+4 Q) \left(m^2+\mathcal{H}^2\right)+a^2 X^2 \mathcal{H}^2 \alpha _{\sigma } \left(Q \left(\mathcal{H}^2-11 m^2\right)-9 J m^2\right)+Q \mathcal{H}^4 \alpha _{\sigma }^2\right] , \quad 
\label{eq:positivity_condition_1}\\
    0 & < & \frac{Q \alpha _{\sigma } \left(\Bar{\chi}'\right)^2 \left(a^2 X^2 \left(m^2+\mathcal{H}^2\right)-\mathcal{H}^2 \alpha _{\sigma }\right)}{a^2 m^2 r X^3 \mathcal{H}^2 M_{\text{Pl}}^2} ,
\label{eq:positivity_condition_2}\\
    0 & < & \frac{\left(\Bar{\chi}'\right)^2}{\mathcal{H}^2 M_{\text{Pl}}^2} . 
\label{eq:positivity_condition_3}
\eea 
\end{widetext}
These inequalities are the Higuchi-type bounds for this specific theory in branch 1. Our bounds are more complicated than the original Higuchi bound \cite{Higuchi:1986py}, $ m^2 \geq 2 H^2 $, or a similar bound for the dRGT theory \cite{Fasiello_2012}, due to the extended structure that now includes $\alpha_\sigma$, the quasidilaton field $\sigma$, and a specific type of matter field. We check whether these conditions can be satisfied in Sec.\ \ref{section:Figures}.

\subsection{Branch $2$: $J = 0$}

In branch $2$, the equations are simple to solve and we do not need to make a subhorizon approximation. We find that the kinetic matrix has {\it two} vanishing eigenvalues rather than one. There is no kinetic term for $\Pi_L$, and its equation of motion thus leads to a constraint. Using the background equation\ (\ref{eq:friedmann2}), specialized to this branch, in fact yields the simple condition that 
\bea
    \delta \sigma \ = \ -\Psi \, .
\label{eq:kmbranch2constraint}
\eea
On substituting this constraint back into the action, using the background equations in this branch, and integrating by parts, the terms with $\Pi_L$ can be completely removed. The final $2 \times 2$ kinetic matrix in the perturbations $(\Psi, m\Pi_0)$ becomes
\bea
    K \, = \, \left(
    \begin{array}{cc}
    -\frac{Q \mathcal{H}^2 \alpha _{\sigma }^2}{a^2 m^2 r^3 X^3}+\frac{\left(\Bar{\chi}'\right)^2}{\mathcal{H}^2 M_{\text{Pl}}^2}+\frac{Q \alpha _{\sigma }}{r X} & \frac{Q \mathcal{H} \alpha _{\sigma } \phi '}{m r^3 X} \\
    \frac{Q \mathcal{H} \alpha _{\sigma } \phi '}{m r^3 X} & \frac{Q \mathcal{H}^2 \alpha _{\sigma }}{m^2 r^3 X} \\
\end{array}
\right) ,
\eea
which leads to the following two positivity conditions, 
\bea 
    \frac{Q \alpha _{\sigma } \left(\Bar{\chi}'\right)^2}{m^2 r^3 X M_{\text{Pl}}^2} & > & 0 \, , \\
    \frac{Q \mathcal{H}^2 \alpha _{\sigma }}{m^2 r^3 X} & > & 0 \, .
\eea
These inequalities are the Higuchi-type bounds in branch 2 and are likely easier to satisfy, but we do not consider them in Sec.\ \ref{section:Figures}, as explained there.


\section{Growth rate}
\label{section:Growth_rate}

In this section, we find how matter perturbations grow in massive gravity. We specifically consider a CDM-dominated Universe, thus restricting to pressureless matter. The only non-vanishing component of the background energy-momentum tensor is then $\bar{T}^0_{0} = -\Bar{\rho}$ and the Friedmann equations describing the background evolution of the Universe are given by Eqs.\ (\ref{eq:friedmann1}) and (\ref{eq:friedmann2}) with $\bar{\rho} = \Omega_{m0}/a^3$, $\Omega_{m0}$ being the total matter density today, and $\bar{p} = 0$. Components of the energy-momentum tensor at first order in the matter perturbation, on the other hand, are given by
\bea
    T^0_{0} &=& -\Bar{\rho}(1+\delta) \, , \\
    T^0_{i} &=& \Bar{\rho}v_i \, , \\
    T^i_{0} &=& -\Bar{\rho}v^i \, , \\
    T^i_{j} &=& 0 \, ,
\eea
where $v^i = dx^i/d\tau$ is the peculiar velocity and is first order in the perturbation. We are interested in the equation of motion for the matter perturbation $\delta(\tau,\vx)$. Unlike the previous section, we find it simplest to work in conformal Newtonian gauge for the calculation here, so that $B = 0$ and $E = 0$; we again show that this is a valid gauge choice in appendix\ \ref{app:gauge}.

Since the energy-momentum tensor is defined with respect to the physical metric, it satisfies the usual conservation equation $\nabla_{\mu} T^{\mu}_{\nu} = 0$. The background piece of the $\nu = 0$ component yields the background continuity equation $\bar{\rho}' + 3{\cal H} \bar{\rho} = 0$, while the first order piece and the $\nu = i$ component yield the first order continuity equation and Euler equation respectively,
\bea
    \delta' - 3\Psi' + \partial_i v^i & = & 0 \, ,
\label{eq:del prime} \\
    v^{i\prime} + \mathcal{H} v^i & = & - \partial^i \Phi \, .
\label{eq:v prime}
\eea
Taking the conformal time derivative of Eq.\ (\ref{eq:del prime}) and using both equations in the result gives
\bea
    \delta'' + \mathcal{H}\delta' - 3\Psi'' - 3 \mathcal{H}\Psi' - \partial^2\Phi & = & 0 \, .
\label{eq:deltapp}
\eea
In Fourier space, the last term above becomes $k^2 \Phi$. We will focus on the linear growth rate in the quasistatic and subhorizon approximations, under which we can take $Y'' \sim {\cal H} Y' \sim {\cal H}^2 Y \ll k^2 Y$ for any perturbation $Y$ in Fourier space and with wavenumber $k$, that are expected to be reasonable approximations \cite{Hojjati:2012rf}; also see \cite{Sawicki:2015zya}. We will also assume that $\Phi$ and $\Psi$ are of a similar order of magnitude. Equation\ (\ref{eq:deltapp}) then becomes
\bea 
    \delta'' + \mathcal{H} \delta' + k^2 \Phi & = & 0 \, .
\label{eq:deltappGR}
\eea
The calculation up till this point is exactly the same as that in GR. To parametrize the departure from GR, we introduce two functions of time (or scale factor) and wave number, $\eta(a,k)$ and $\mu(a,k)$, following the notation of \cite{Baker:2019gxo},
\bea
    \Phi & = & \eta(a,k)\Psi \, ,
\label{eq:eta_definition} \\
    k^2 \Psi & = & -4\pi G a^2 \mu(a,k)\Bar{\rho}\delta \, .
\label{eq:mu_definition}
\eea
Both $\eta$ and $\mu$ reduce to unity in GR. We are now left with obtaining and solving the equations of motion for $\eta$ and $\mu$ in terms of background quantities. As shown below, we find that both are in fact independent of $k$ under the approximations that we have made, consistent with the generalized massive gravity \cite{Kenna-Allison:2020egn}, projected massive gravity \cite{Manita:2022}, and minimal theory of massive gravity \cite{DeFelice:2021trp} frameworks. This is in contrast to what one finds in bi-metric massive gravity \cite{Solomon:2014dua} and, for example, $f(R)$ theories of gravity \cite{Zhang:2005vt,Bean:2006up}. 

In order to derive the equations of motion for various perturbations, we first obtain the second-order action and then set its variation to zero. This is similar to how we obtained the background equations of motion by setting $\delta S^{(0)}$ to zero in Eq.\ (\ref{eq:varying action}), except that we now set $\delta S^{(2)}$ to zero, with the variation taken with respect to all first-order perturbations. Note that while writing the second-order matter action $S_{\rm matter}^{(2)}$ can be tricky, its variation can be obtained directly from the definition of the energy-momentum tensor in Eq.\ (\ref{eq:stress-energy tensor}),
\bea
    \delta S_\mathrm{matter} & = & \int d^4 x \, \frac{\delta}{\delta g^{\mu\nu}} \( \sqrt{-g} \, \mathcal{L}_\mathrm{matter} \) \delta g^{\mu\nu} \nn \\
    & = & \frac{1}{2} \int d^4 x \sqrt{-g} \, T^\mu_\lambda g^{\lambda\nu} \delta g_{\mu\nu} \, ,
\label{eq:deltaSm}
\eea
where we used $\delta g^{\mu\nu} = -g^{\mu\rho}g^{\nu\sigma}\delta g_{\rho\sigma}$. The variation $\delta g_{\mu\nu}$ is first order in the perturbations and contains the {\it variation} of metric perturbations: $\delta\Phi$, $\delta B$, $\delta\Psi$, and $\delta E$. Expanding out all other pieces in Eq.\ (\ref{eq:deltaSm}) to first order in the perturbations then yields the variation of the second-order matter action $\delta S_\mathrm{matter}^{(2)}$.

We do not show explicit expressions for $\delta S^{(2)}$ and the equations of motion before approximations as they are quite big, but they can be found in our {\tt Mathematica} supplement. In the following two subsections, we analyze how matter perturbations grow under the quasistatic and subhorizon approximations in each of the two branches of solutions found earlier.

\subsection{Branch 1: $\sigma' = \mathcal{H}r$}

We find it simpler to impose the conformal Newtonian gauge condition ($B = 0$ and $E = 0$) at the level of the equations of motion rather than the action. This allows us to obtain an equation for $E$, equivalent to the $ij \ (i \neq j)$ component of the field equations, which informs us of the scaling of $\Pi_L$. We find that the $E$ equation of motion, before imposing the branch condition or making any approximations, but after setting the gauge condition, is
\begin{widetext}
\bea 
    \Phi - \Psi + \frac{a^2 m^2 X\left(J (X-1) (r (X+1)-2)-(r-1) \left(Q-X^3+X^2\right)\right)}{(X-1)^2} \Pi_L & = & 0 \, .
\label{eq:E_EoM}
\eea 
\end{widetext}
This suggests that the dimensionful perturbation $\Pi_L$ scales as $m^2 \Pi_L \sim \Phi, \, \Psi$. We similarly use $m$ to scale $\Pi_0$, so that $m \Pi_0 \sim \Phi, \, \Psi$ as well. Lastly, we assume that $\delta\sigma$ is also of similar order as $\Phi$ and $\Psi$, thus ending up with the scaling used earlier in Sec.\ \ref{section:Kinetic_matrix}.

With these scaling relationships and assuming that $ma \lesssim {\cal H}$, the $\Phi$ equation of motion in branch 1, under the quasistatic and subhorizon approximations, and after setting the gauge condition, is
\bea 
    k^2 \left(J m^2 X \Pi _L-\frac{2 \Psi }{a^2}\right)-\frac{  \bar{\rho }}{M_{\text{Pl}}^2}\delta & = & 0 \, .
\label{eq:Phi_EoM}
\eea
Under the same set of conditions, the equations of motion for $\delta\sigma$ and $\Pi_0$, after substituting the background equations of motion, are given respectively by
\begin{widetext}
\bea 
    0 & = & \left(a^2 m^2 \left(J^2 (X-1)^2 \alpha _{\sigma } (r X^2-2 X+1)+J (X-1) \left(Q (X-1) \alpha _{\sigma } (r X^2-2 X+2)+2 X^3 \left(r (2-6 X)+3 X+1\right)\right.\right.\right. \nn\\
    & & -X (X-1)^3 \left.\left.\alpha _{\sigma }\right)+Q^2 (X-1)^2 \alpha _{\sigma }-Q X \left(4 X^2 \left(r (2 X^2-4 X+1)+1\right)+(X-1)^4 \alpha _{\sigma }\right)-4 (r-1) (X-1) X^5\right) \nn \\
    & & +X \left.\mathcal{H}^2 \alpha _{\sigma } \left(J (X-1) \left(2 r (5 X-1)-5 X-3\right)+Q \left(2 r (3 X^2-6 X+1)-X^2+2 X+3\right)+4 (r-1) (X-1) X^2\right)\right) \nn \\
    & & \frac{1}{2 (X-1)^2 X \alpha _{\sigma }}\Pi _L  + \left(\frac{r \mathcal{H}^2 \alpha _{\sigma } \left(J r+Q (r+1)\right)}{a^2 m^2 (r+1) X^2}-J (r-1)-Q r\right)\delta \sigma \nn \\
    & & +\frac{J \mathcal{H}}{r+1}\Pi _L'+\frac{ \mathcal{H} \phi ' (J r+Q (r+1))}{r+1}\Pi _0 \, ,
\eea
\bea 
    0 & = & -\frac{a^2 m^2 (r-1) X^2 \mathcal{H} \phi' \left(3 J^2 (X-1)^2+J Q \left(X^2-6 X+5\right)+2 Q \left(Q-X^3+X^2\right)\right)}{Q (X-1)^2} \Pi _L +\frac{a^2 J m^2 X^2 \phi ' }{r+1}\Pi _L' \nn \\
    & & +\frac{r \left(a^4 J m^2 X^2-a^2 r \mathcal{H}^2 \alpha _{\sigma } (J r+Q (r+1))\right)}{a^2 (r+1)}\Pi_0 +\frac{  r \mathcal{H} \alpha _{\sigma } \phi' (J r+Q (r+1))}{r+1}\delta \sigma \, .
\eea
\end{widetext}
These two equations can be solved simultaneously for $\delta\sigma$ and $\Pi_0$ to obtain solutions of the form
\bea
    \delta\sigma & = & c_1 \Pi_L -\mathcal{H} \Pi_L' \, ,
\label{eq:deltasigma_sol} \\
    \Pi_0 & = & c_2 \Pi_L -\frac{\phi '}{r} \Pi_L' \, ,
\label{eq:Pi_0_sol}
\eea
where the coefficients $c_1$ and $c_2$ are functions of background quantities. Lastly, the $\Pi_L$ equation of motion under the same set of conditions is given by
\begin{widetext}
\bea 
    0 &=& \left[Q r \left(\alpha _{\sigma } \left((r+1) \left(a^2 m^2 (X-1)^2 \left(J^2 r \left(r X^2-2 X+1\right)+J Q \left(r^2 X^2+r \left(X^2-2 X+2\right)-2 X+1\right) \right.\right.\right.\right.\right.\nn\\
    && \left.-J r (X-1)^2 X+Q (r+1) \left(Q-(X-1)^2 X\right)\right)+X \mathcal{H}^2 \left(J (X-1) \left(2 r^2 (5 X-1)+r (3 X-11)-6 X+6\right) \right.\nn\\
    && \left.\left.\left.+Q \left(2 r^2 \left(3 X^2-6 X+1\right)+r \left(5 X^2-10 X+9\right)-(X-1)^2\right)+4 (r-1) r (X-1) X^2\right)\right)+2 J X (X-1)^2 \mathcal{H} r'\right) \nn\\
    && \left.\left.-2 a^2 m^2 (r+1)^2 X^3 \left(J (X-1) (r (6 X-2)-3 X-1)+2 \left(Q \left(2 r X^2-4 r X+r+1\right)+(r-1) (X-1) X^2\right)\right)\right)\right]\delta\sigma \nn\\
    && +2 a^2 m^2 X^3 \phi ' \left[J Q (X-1)^2 r'-r \left(r^2-1\right) \mathcal{H} \left(3 J^2 (X-1)^2+J Q \left(X^2-6 X+5\right)+2 Q \left(Q-X^3+X^2\right)\right)\right]\Pi_0 \nn\\
    && +2 a^2 m^2 Q r X^3 \left[J (X-1)^2 r'-(r+1) \mathcal{H} \left(J (X-1) (3 r X+r+X-5)-2 (r-1) \left(Q-X^3+X^2\right)\right)\right]\Pi _L' \nn\\
    && -2 a^2 J m^2 Q r (r+1) (X-1)^2 X^3 \Pi _L''-2 a^2 J m^2 Q (r+1) (X-1)^2 X^3  \phi '\Pi _0'+2 a^2 J m^2 Q r (r+1)^2  (X-1)^2 X^3\Phi  \nn\\
    && -4 a^2 m^2 Q r (r+1)^2   X^3 \left(J (X-1) (r X+r-2)-(r-1) \left(Q-X^3+X^2\right)\right)\Psi \nn\\
    && -2 J Q r (r+1) (X-1)^2 X \mathcal{H} \alpha _{\sigma } \delta \sigma' \, .
\label{eq:Pi_L_EoM}
\eea 
\end{widetext}
We now have five equations in six perturbations: $\Phi$, $\Psi$, $\delta\sigma$, $\Pi_0$, $\Pi_L$, and $\delta$. We next solve them to obtain the functions $\eta$ and $\mu$ defined in Eqs.\ (\ref{eq:eta_definition}) and (\ref{eq:mu_definition}).

We first take a conformal time derivative of $\delta\sigma$ and $\Pi_0$ in Eqs.\ (\ref{eq:deltasigma_sol}) and (\ref{eq:Pi_0_sol}), and substitute for both perturbations and their conformal time derivatives into the $\Pi_L$ equation of motion in Eq.\ (\ref{eq:Pi_L_EoM}). On doing so, the terms with $\Pi_L''$ cancel out. We next eliminate $\Pi_L'$ by taking a time derivative of the $E$ equation of motion in Eq.\ (\ref{eq:E_EoM}) and substituting it into Eq.\ (\ref{eq:Pi_L_EoM}) as well. In the quasistatic and subhorizon approximations, the resulting $\Pi_L$ equation of motion then becomes a linear equation of the form $\Phi = \eta(a,k) \Psi$. Finally, using the $\Phi$ equation of motion in Eq.\ (\ref{eq:Phi_EoM}), gives us a modified Poisson equation $k^2 \Psi = -4\pi G a^2 \mu(a,k)\Bar{\rho}\delta$. The resulting expressions for $\eta$ and $\mu$ can be written in a simple form as
\bea
    \eta & = & \frac{c_{\eta 1}a^4 m^2 -c_{\eta\mu} \mathcal{H}^2}{c_{\eta 2}a^4 m^2 - c_{\eta\mu} \mathcal{H}^2} \xrightarrow[]{m\rightarrow 0} 1 \, ,
\label{eq:eta} \\
    \mu & = & \frac{c_{\mu 1}a^4 m^2 -c_{\eta\mu} \mathcal{H}^2}{c_{\mu 2}a^4 m^2 - c_{\eta\mu} \mathcal{H}^2} \xrightarrow[]{m\rightarrow 0} 1 \, .
\label{eq:mu}
\eea
Since this is one of our main results, we show explicit expressions for the coefficients \{$c_{\eta 1}, c_{\eta 2}, c_{\eta\mu}, c_{\mu 1}, c_{\mu 2}$\} in appendix\ \ref{app:full_expressions}. As can be seen from these equations, $\eta$ and $\mu$ are both independent of $k$.

The resulting expression for $\Phi$ can now be plugged into Eq.\ (\ref{eq:deltappGR}) to obtain a second-order differential equation in the matter perturbation $\delta$,
\bea 
    \delta'' + \mathcal{H}\delta' - 4\pi G a^2\mu\eta \Bar{\rho}\delta  & = & 0 \, .
\eea
Making use of the fact that $\eta$ and $\mu$ are independent of $k$, the above equation suggests that $\delta$ can be factorized as $\delta(\tau, \vk) = D(\tau) \delta(\vk)$, which is similar to what one finds in GR, with $D(\tau)$ being the growth factor. The evolution of $D(\tau)$ is described by
\bea 
    D'' + \mathcal{H}D' - 4\pi G a^2\mu\eta \Bar{\rho}D & = & 0 \, ,
\eea
and directly gives us the growth rate as well using $f = d \ln D/d \ln a$. We solve for $\eta$, $\mu$, $D$, and $f$ numerically in the next section.

\subsection{Branch 2: $J=0$}

On using the branch 2 condition that $J = 0$ and the background equations in this branch, the $\Pi_L$ equation of motion simplifies significantly to give
\bea
    \delta \sigma & = & -\Psi \, ,
\eea
in agreement with what we found in the previous section in Eq.\ (\ref{eq:kmbranch2constraint}). From the gauge transformations in appendix\ \ref{app:gauge}, along with the background constraint that $\sigma' = {\cal H}$ in branch 2, we see that the combination $\delta\sigma + \Psi - (1/3) k^2 E$ is gauge-invariant in this branch. Since we set $E = 0$ in both gauge choices that we have made, that in the previous section and in the current section, it is not surprising that we find the same $\delta\sigma + \Psi$ in both cases.

Now on substituting $\delta\sigma = -\Psi$ into the $\delta\sigma$ and $\Pi_0$ equations of motion in this branch, we can solve them simultaneously to find that $\Pi_L$ vanishes. Using this, and that $J = 0$, in turn in Eq.\ (\ref{eq:E_EoM}), that was valid in both branches, yields $\Phi = \Psi$. Note that this coincides with what one finds in GR. Lastly, substituting $\Phi = \Psi$ in the $\Phi$ equation of motion in this branch and working under the quasistatic and subhorizon approximations gives us the Poisson equation $k^2 \Psi = -4\pi G a^2 \Bar{\rho}\delta$, which again coincides with what one finds in GR. 

Both $\eta$ and $\mu$ are thus unity in this branch and there is no departure from GR at the level of first-order perturbations. Further, the differential equation for the matter perturbation $\delta$ also coincides with that in GR,
\bea 
    \delta'' + \mathcal{H}\delta' - 4\pi G a^2 \Bar{\rho}\delta  & = & 0 \, .
\eea
The growth factor $D$ satisfies the same equation, as noted before. Note, however, that at the background level, the Friedmann equations do contain contributions from massive gravity.


\section{Numerical solutions}
\label{section:Figures}

We are next interested in solving the equations obtained in the previous sections to understand whether the results can match cosmological observations. Our goal here is not to perform a detailed Markov-Chain Monte-Carlo (MCMC) analysis over the parameters of the theory, but rather to demonstrate how various observables evolve in massive gravity, compared to a $\Lambda$CDM Universe. We therefore choose a set of parameter values that provides a reasonable fit to the data solely for illustration; best-fit parameter values would likely be different. We further restrict to branch 1 for the analysis in this section. This choice is motivated by the fact that, in the absence of matter, vector modes in branch 2 become infinitely strongly coupled \cite{Gumrukcuoglu:2017ioy}, although we have not checked whether this issue persists in the presence of matter. If branch 2 can be stabilized, then, as shown in Sec.\ \ref{section:Growth_rate}, the growth factor evolution could match that in standard cosmology with the massive gravity piece in the Friedmann equation\ (\ref{eq:friedmann1}) acting as an effective cosmological constant.

We first perform a search over the four parameters of the theory $\{ m, \alpha_3, \alpha_4, \alpha_{\sigma} \}$ and $\Omega_{m0}$. For a given set of parameter values, we choose $X$ today, that we denote $X_0$, by solving the Friedmann equation\ (\ref{eq:friedmann1}) with the Hubble parameter set to $h = 0.6751$, consistent with the Planck 2015 TT,TE,EE$+$lowP$+$lensing result \cite{Planck:2015fie}. We set the radiation density today to zero, which is a good approximation at the redshifts that we are interested in. This yields three roots for $X_0$, of which we choose any one to proceed with. With this initial condition, we evolve $X$ by solving an equation in $dX/d\ln a$ that we obtain by combining the background equation for $\sigma$, Eq.\ (\ref{eq: sigma background}), with the branch 1 condition, $\sigma' = \mathcal{H}r$, and the evolution equation $X' = (r-1) {\cal H}X$ for $X$, which gives
\begin{widetext}
\bea 
    \frac{dX}{d\ln a} = -\frac{Q \left(a^4 m^2 \left(J (X-1)^2 \alpha _{\sigma }+Q \alpha _{\sigma }-X \left(8 X^2+(X-1)^2 \alpha _{\sigma }\right)\right)+5 a^2 X \mathcal{H}^2 \alpha _{\sigma }\right)}{a^4 m^2 X \left(J \left(Q \alpha _{\sigma }-6 X\right)-8 Q X\right)+6 a^2 \mathcal{H}^2 \alpha _{\sigma } (J+Q)} \, .
\eea 
\end{widetext}
We then demand that the evolution of $X$ remains smooth up to a redshift of $z = 30$ so that $X$ does not jump between different roots, discarding any set of parameter values that do not satisfy this condition.

We also demand that the positivity conditions in Eqs.\ (\ref{eq:positivity_condition_1})-(\ref{eq:positivity_condition_3}) are satisfied up to $z = 30$, although these conditions were derived for a scalar field matter component rather than CDM. To do so, we first need to specify a potential $V(\bar{\chi})$ and find $\bar{\chi}'^2$ by numerically solving the Klein-Gordon equation,
\bea
    \bar{\chi}'' + 2\mathcal{H} \bar{\chi}' + a^2 \partial_{\bar{\chi}} V(\bar{\chi}) & = & 0 \, .
\eea
Since our end goal is to compare observables in massive gravity with those in $\Lambda$CDM, we are specifically interested in the case that the scalar field $\bar{\chi}$ behaves like dark matter (see, for example, \cite{Magana:2012}, \cite{Urena-Lopez:2019kud} for a review on scalar field dark matter models). We consider a potential of the form $V(\bar{\chi}) = (1/2) m_{\chi}^2 \bar{\chi}^2$ in the limit that $m_{\chi} \gg \mathcal{H}$, where the oscillation period of the field $\bar{\chi}$ is much smaller than the rate of Hubble expansion. $\bar{\chi}'^2$ can then be replaced by its average value over the oscillation period, in which case the field $\bar{\chi}$ behaves like non-relativistic matter, with vanishing average pressure \cite{Turner:1983he}.\footnote{Also see \cite{Gumrukcuoglu:2016hic} for a discussion of issues that may arise in the pressureless scalar field scenario.} In this regime, the average value of $\bar{\chi}'^2$ equals the energy density $\bar{\rho}$. We can thus simply solve the first Friedmann equation\ (\ref{eq:friedmann1}) for $\bar{\chi}'^2$, which we then substitute in the positivity conditions. We again discard any set of parameter values that does not satisfy the resulting positivity conditions at any time between $z = 30$ and today. Lastly, we also demand that the functions $\eta$ and $\mu$ given in Eqs.\ (\ref{eq:eta}) and (\ref{eq:mu}) remain positive between $z = 30$ and today.

For a set of parameter values that satisfies the above constraints, the solution for $X(z)$ allows us to find the Hubble parameter at any redshift, which we compare against $580$ distance modulus measurements from the Supernova Cosmology Project \cite{Suzuki:2011hu}. We also calculate how the growth rate evolves in branch 1, setting the initial condition $D(a) = a$ for the growth factor deep within the matter dominated era, at a redshift of $z = 30$, and compare it against $30$ growth rate measurements compiled in, for example, \cite{Perenon:2019dpc}. We demand that the $\chi^2$ of the fits to these two data sets is less than a thousand, and choose one set of parameter values that satisfies this condition to make the figures below. The values that we use are $m/H_0 = 10^{-1.410}$, $\alpha_3 = -3.265$, $\alpha_4 = 3.267$, $\alpha_{\sigma} = 10^{-0.4200}$, and $\Omega_{m0} = 0.3721$, with $X_0 = 14.30$. 
These yield a $\chi^2$ of approximately $805$. For comparison, a two-parameter fit over $\{ \Omega_{m0}, \Omega_{\Lambda 0} \}$ for a $\Lambda$CDM cosmology within GR gives a best-fit $\chi^2 \approx 579$ with the same datasets. We note again that the parameter values that we choose are {\it not} the best-fit parameter values that would result from an MCMC analysis.

\begin{figure}[!t]
\begin{center}
	\includegraphics[width=3in,angle=0]{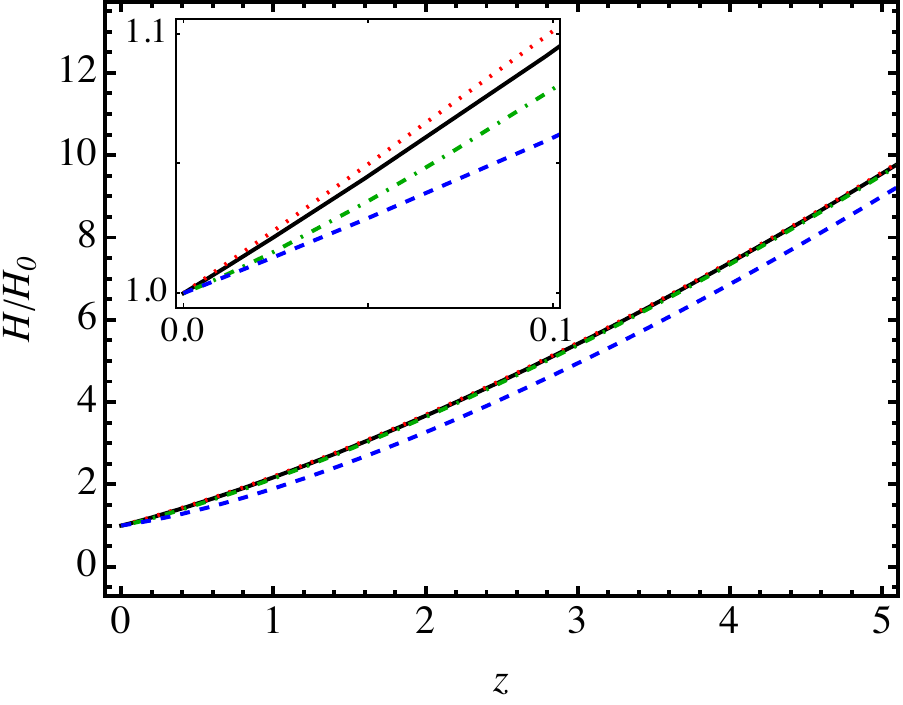}
	\caption{The Hubble parameter $H(z)/H_0$ in massive gravity (black, solid) and $\Lambda$CDM (blue, dashed) for the parameter values mentioned in the text. Also shown are graphs for $m/H_0$ being $0.5$ times its central value (red, dotted) and $1.5$ times its central value (green, dot-dashed). The three massive gravity curves are almost indistinguishable, and the inset zooms into a part of the figure to distinguish the different curves.}
\label{fig:H_plot}
\end{center}
\end{figure}

\begin{figure*}[!t]
\begin{center}
	\includegraphics[width=3.09in,angle=0]{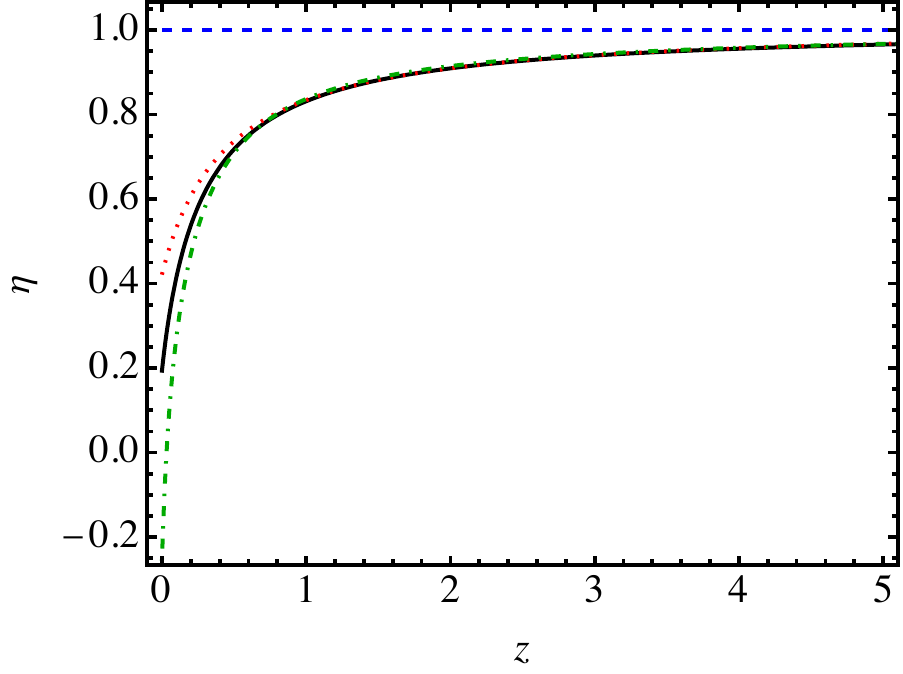}
	\includegraphics[width=3in,angle=0]{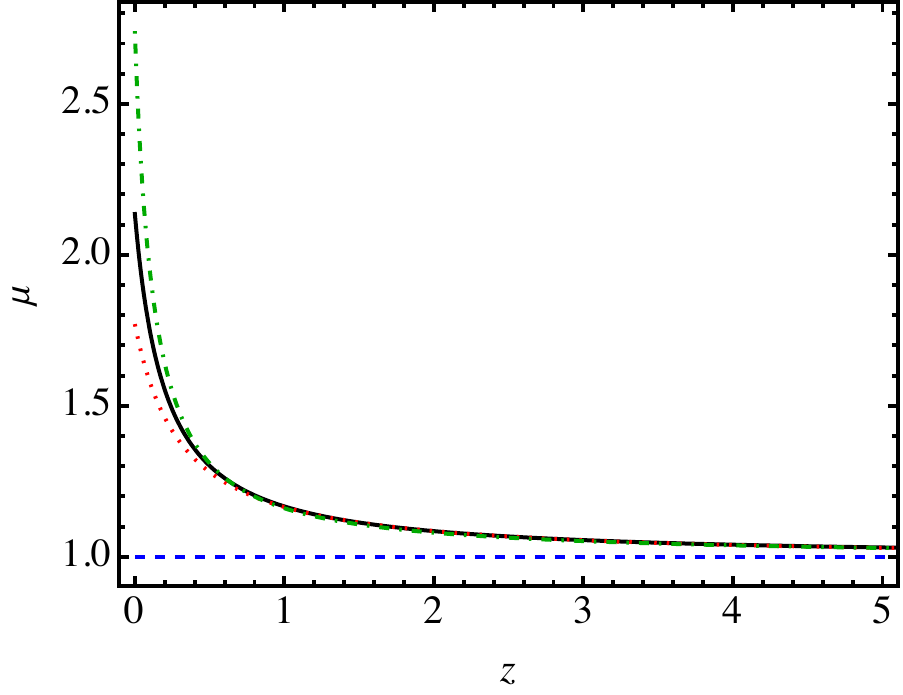}
	\caption{The functions $\eta(z)$ (left panel) and $\mu(z)$ (right panel) in massive gravity (black, solid) and $\Lambda$CDM (blue, dashed) for the parameter values mentioned in the text. Also shown are graphs for $m/H_0$ being $0.5$ times its central value (red, dotted) and $1.5$ times its central value (green, dot-dashed).}
\label{fig:eta_mu_plot}
\end{center}
\end{figure*}

\begin{figure*}[!t]
\begin{center}
	\includegraphics[width=3in,angle=0]{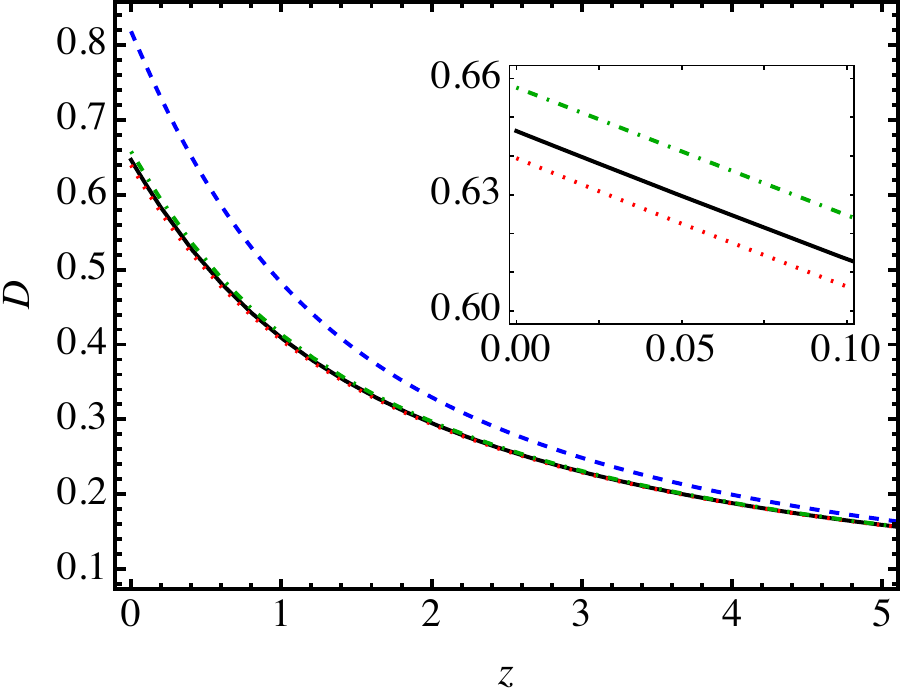}
	\includegraphics[width=3in,angle=0]{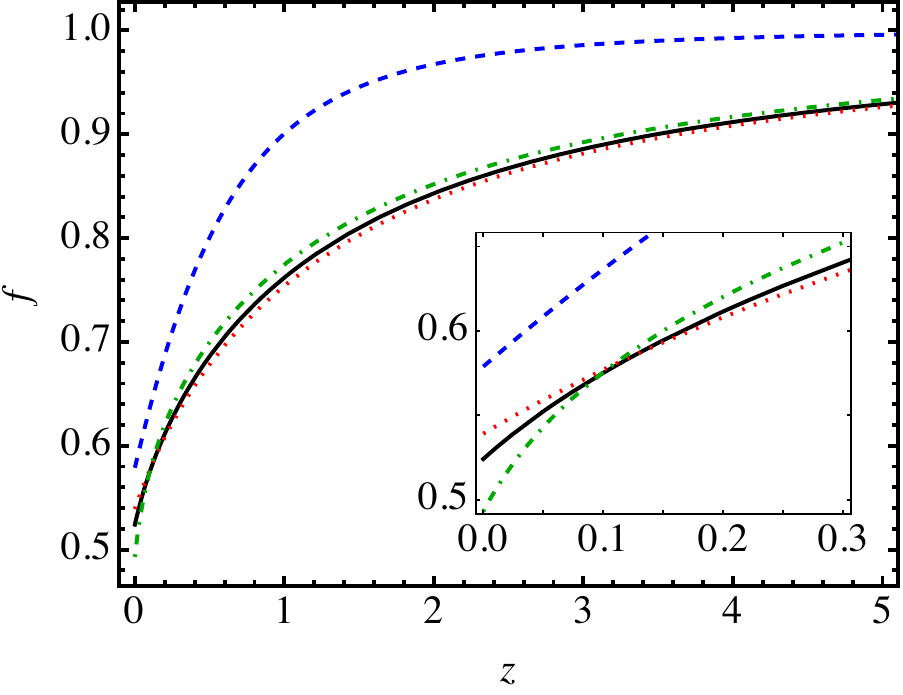}
	\caption{The linear growth factor $D(z)$ (left panel) and growth rate of structure $f(z)$ (right panel) in massive gravity (black, solid) and $\Lambda$CDM (blue, dashed) for the parameter values mentioned in the text. Also shown are graphs for $m/H_0$ being $0.5$ times its central value (red, dotted) and $1.5$ times its central value (green, dot-dashed). The insets zoom into a part of the figure to distinguish the different curves.}
\label{fig:D_f_plot}
\end{center}
\end{figure*}

For the parameter values mentioned above, we show how the Hubble parameter $H(z)$ evolves in massive gravity compared to $\Lambda$CDM in Fig.\ \ref{fig:H_plot}. We also show $H(z)$ for two slightly different values of $m/H_0$, $0.5 \times 10^{-1.410}$ and $1.5 \times 10^{-1.410}$, keeping all other parameters except for $X_0$ fixed to their previous values. For $X_0$, we solve the Friedmann equation again so that the Hubble parameter today is still $H_0$. In Fig.\ \ref{fig:eta_mu_plot}, we show the evolution of the functions $\eta(z)$ and $\mu(z)$, both of which, as noted earlier, are independent of the wave number under the quasistatic and subhorizon approximations, and are unity in GR. Lastly, we show how the linear growth factor $D(z)$ and growth rate $f(z)$ evolve in massive gravity and $\Lambda$CDM in Fig.\ \ref{fig:D_f_plot}. The figures show good agreement with the evolution of $H(z)$ in $\Lambda$CDM but differences with $\eta(z)$, $\mu(z)$, $D(z)$, and $f(z)$ at late times. We find a similar evolution for all five observables when we vary $\alpha_3$, $\alpha_4$, or $\alpha_\sigma$ instead, while keeping other parameters (except for $X_0$) fixed to their previous values. We also need to keep $\alpha_3 \approx -\alpha_4$, however, which we find is required for $\eta(z)$ and $\mu(z)$ to not diverge.

There are two subtle points worth mentioning before we close this section. First, we found that the positivity condition in Eq.\ (\ref{eq:positivity_condition_1}) is sensitive to the form of the potential $V(\bar{\chi})$. For example, if we consider a potential in the limit that $\bar{\chi}'^2 \gg V(\bar{\chi})$, then the field behaves like a fluid with $\bar{p} = \bar{\rho}$, and the expression on the right-hand side in Eq.\ (\ref{eq:positivity_condition_1}) becomes negative. In general, the condition in Eq.\ (\ref{eq:positivity_condition_1}) seems to be satisfied only when we fine-tune parameter values and/or the potential such that the $\bar{\chi}'^4$ term is smaller than the $\bar{\chi}'^2$ one. This is in agreement with the suggestion that the theory is pathological in the presence of matter, leading either to an instability or decelerated expansion.\footnote{We thank Emir G\"{u}mr\"{u}k\c{c}\"{u}o\u{g}lu for pointing out that even in the presence of a cosmological constant, one finds either an instability or that the cosmological constant must be negative.} We note, however, that a pressureless scalar field, as considered earlier in this section, appears to resolve the problem.

Second, we found that for values of $\Omega_{m0}$ close to its observed mean value of $0.3121$ \cite{Planck:2015fie} and for certain choices of other parameter values, $\mu$ diverges at some time between $z=30$ and $z=0$ . This arises from a vanishing denominator of $\mu$, which Eq.\ (\ref{eq:mu_definition}) suggests is indicative of the matter overdensity $\delta$ passing through zero. A similar zeroing of $\delta$ was observed for $f(R)$ gravity in \cite{Zhang:2005vt}, and was attributed to the quasistatic approximation in \cite{Bean:2006up}. We thus suspect that the quasistatic approximation may be the reason for the divergence of $\mu$ that we find for some sets of parameter values as well. We resolved this issue by allowing $\Omega_{m0}$ to be a free parameter in our parameter search, and in fact found that a slightly higher matter content (we used $\Omega_{m0} = 0.3721$ as mentioned earlier) prevented $\delta$ from passing through zero.


\section{Summary and discussion}
\label{section:Discussion}

Growth of structure is a promising probe of GR and its modifications. In this paper, we were interested in the growth of CDM perturbations in a model of massive gravity that allowed for a self-accelerated solution without the need for a cosmological constant and was stable under scalar perturbations. We specifically considered the extended quasidilaton setup of massive gravity without a kinetic term for the quasidilaton field, that was shown to admit a stable solution in the absence of matter in \cite{Gumrukcuoglu:2017ioy}. We first obtained the background equations in the presence of matter and found two branches of solutions, denoted branches 1 and 2, in agreement with what one finds in the absence of matter. We next considered two scalar perturbation calculations, one with matter modeled as a scalar field and one with CDM, that we summarize below.

In order to check whether scalar perturbations are stable in the presence of matter, we considered a scalar field matter component. We obtained the second-order action while working in unitary gauge for the matter field and, after integrating out non-propagating degrees of freedom, found the kinetic matrix in the remaining perturbations. In branch 1, demanding that the resulting kinetic matrix is positive definite led to three conditions. By performing a parameter search, we found that one of the three conditions is not satisfied in general, but is satisfied for a particular choice of potential where the scalar field effectively behaves like pressureless matter. In branch 2, on the other hand, we found two conditions, which are likely easier to satisfy, but we did not consider in detail since vector perturbations become infinitely strongly coupled in the absence of matter in this branch \cite{Gumrukcuoglu:2017ioy}.

We next studied the growth of CDM perturbations and obtained the equations of motion for various perturbations while working in conformal Newtonian gauge. We obtained expressions for the observables $\eta$ and $\mu$, that are commonly used to parametrize any deviation from GR, and equations for the growth factor and growth rate in both branches. Focusing again on branch 1, and for a choice of parameter values and potential that lead to a positive definite kinetic matrix and provide a reasonable fit to background and growth of structure data, we showed how the background Hubble parameter and perturbations evolve between $z = 0$ and $z = 5$, compared to $\Lambda$CDM. The parameter values that we chose were not the best-fit values resulting from a full MCMC analysis, but we expect the evolution of observables to be qualitatively similar for other parameter values.

We found, for example, that the Hubble parameter can be close to that in $\Lambda$CDM, even though we have not added a cosmological constant. We also found that $\eta$ and $\mu$ are independent of $k$ in the quasistatic and subhorizon approximations, which is in contrast with $f(R)$ theories of gravity, where one finds that both $\eta$ and $\mu$ are scale-dependent under the subhorizon and quasistatic approximations \cite{Zhang:2005vt,Bean:2006up}. Therefore, while observing a $k$-dependence would rule out GR, not observing one would not immediately rule out massive gravity. Both observables, however, do differ from their GR values of unity at late times, and smoothly go to unity at high redshifts. The growth factor and growth rate similarly show deviations from their $\Lambda$CDM values at late times. By varying the graviton mass slightly around the central value that we considered, we further found that all observables only shift by a small amount, and it may thus be hard to find parameters that match the evolution of perturbations in $\Lambda$CDM.

In conclusion, the model of massive gravity that we studied can potentially agree with the background expansion in $\Lambda$CDM and background data, but shows deviations from $\Lambda$CDM in the growth of matter perturbations at late times. Scalar perturbations are further stable only for fine-tuned parameter values and a specific choice of potential. Alternative mechanisms for stabilizing perturbations to the St\"{u}ckelberg fields may thus be a promising direction to  pursue in the future. Recent investigations include, for example, a generalized matter coupling \cite{Gumrukcuoglu:2019rsw}, generalized massive gravity \cite{Kenna-Allison:2020egn}, projected massive gravity \cite{Manita:2022}, the minimal theory of bigravity \cite{DeFelice:2020ecp}, massive gravity with non-minimal coupling \cite{Gumrukcuoglu:2020utx}, Gauss-Bonnet massive gravity \cite{Akbarieh:2021vhv}, and the minimal theory of massive gravity \cite{DeFelice:2021trp}.



\acknowledgments

It is a pleasure to thank Amin Akbarieh, Emir G\"{u}mr\"{u}k\c{c}\"{u}o\u{g}lu, Yousef Izadi, Tina Kahniashvili, and George Lavrelashvili for useful discussions. A.~M. was supported in part through a summer fellowship awarded by the Kennedy College of Sciences at UML. A.~R.~P. was supported by NASA under award numbers 80NSSC18K1014, NNH17ZDA001N, and 80NSSC22K0666, and by the NSF under award number 2108411. A.~R.~P. was also supported by the Simons Foundation.


\appendix
\section{Choice of gauge}
\label{app:gauge}

\renewcommand{\theequation}{A\arabic{equation}}
\setcounter{equation}{0}

In this appendix, we construct two sets of gauge-invariant scalar perturbations, to show that non-zero perturbations coincide with gauge-invariant ones in each of the two gauge choices made in the main text, in Secs.\ \ref{section:Kinetic_matrix} and \ref{section:Growth_rate}. We will consider an infinitesimal coordinate transformation of the form $x^\alpha \rightarrow \tilde{x}^\alpha(t,\vx) = x^\alpha + \xi^\alpha(t,\vx)$, with $\xi^\mu = (\xi^0, \xi^i + \partial^i\xi_L)$, $\xi^0$ and $\xi_L$ being scalars. $\xi^i$ is the vector part of the coordinate transformation that we ignore in this paper. Under this, and going to Fourier space, the four scalar perturbations in the dynamical metric, two in the St\"{u}ckelberg fields, and one in the quasidilaton field transform as

\bea
    \widetilde{\Phi} & = & \Phi - \frac{1}{a} (a\xi^0)' \, , \\
    \widetilde{B} & = & B + \xi^0 - \xi_L' \, , \\
    \widetilde{\Psi} & = & \Psi + \mathcal{H} \xi^0 -\frac{1}{3} k^2 \xi_L \, , \\
    \widetilde{E} & = & E - \xi_L \, , \\
    \widetilde{\Pi}^0 & = & \Pi^0 -\phi' \xi^0 \, , \\
    \widetilde{\Pi}_L & = & \Pi_L -  \xi_L \, , \\
    \widetilde{\delta\sigma} & = & \delta\sigma - \sigma' \xi^0 \, .
\eea
Lastly, the perturbation $\delta\chi(\tau,\vk)$ in a matter component consisting of a scalar field $\chi$ and the CDM perturbation $\delta(\tau,\vk)$ transform as
\bea
    \widetilde{\delta\chi} & = & \delta\chi - \frac{a^2 \bar{\rho}(1+w)}{\bar{\chi}'} \xi^0 \, , \\
    \widetilde{\delta} & = & \delta + 3\mathcal{H} \xi^0 \, ,
\eea
where $w = \bar{p}/\bar{\rho}$ is the equation of state for the scalar field and we have used the background continuity equation to simplify the second equation above.

With the transformation equations in hand, we can construct different sets of gauge-invariant perturbations. The first set that we consider is obtained by solving for $\xi^0$ from the $\widetilde{\delta\chi}$ equation and $\xi_L$ from the $\widetilde{E}$ one, which leads to the following gauge-invariant quantities,
\bea
    \Phi_{\rm G.I.} & = & \Phi -\frac{1}{a} \[ \frac{\bar{\chi}'}{a\bar{\rho}(1+w)} \delta\chi \]' , \\
    B_{\rm G.I.} & = & B + \frac{\bar{\chi}'}{a^2 \bar{\rho}(1+w)} \delta\chi - E' \, , \\
    \Psi_{\rm G.I.} & = & \Psi + \frac{\mathcal{H} \bar{\chi}'}{a^2 \bar{\rho}(1+w)} \delta\chi - \frac{1}{3} k^2 E \, , \\
    \Pi^0_{\rm G.I.} & = & \Pi^0 - \frac{\phi' \bar{\chi}'}{a^2 \bar{\rho}(1+w)} \delta\chi \, , \\
    \Pi_{L,{\rm G.I.}} & = & \Pi_L - E \, , \\
    \delta\sigma_{\rm G.I.} & = & \delta\sigma - \frac{\sigma' \bar{\chi}'}{a^2 \bar{\rho} (1+w)} \delta\chi \, .
\eea
Under the gauge choice $E = 0$ and $\delta\chi = 0$ that we made in Sec.\ \ref{section:Kinetic_matrix}, therefore, the perturbations $\Phi$, $B$, $\Psi$, $\Pi^0$, $\Pi_L$, and $\delta\sigma$ coincide with their gauge-invariant counterparts.

The second set of gauge-invariant perturbations that we consider is obtained by solving for $\xi^0$ from the $\widetilde{B}$ equation and $\xi_L$ from the $\widetilde{E}$ one, which gives
\bea
    \Phi_{\rm G.I.} & = & \Phi + \frac{1}{a} \[ a(B-E') \]' \, , \\
    \Psi_{\rm G.I.} & = & \Psi - \mathcal{H}(B-E') - \frac{1}{3} k^2 E \, , \\
    \Pi^0_{\rm G.I.} & = & \Pi^0 + \phi' (B-E') \, , \\
    \Pi_{L,{\rm G.I.}} & = & \Pi_L -E \, , \\
    \delta\sigma_{\rm G.I.} & = & \delta\sigma + \sigma' (B-E') \, , \\
    \delta_{\rm G.I.} & = & \delta - 3\mathcal{H} (B-E') \, .
\eea
Under the gauge choice $B = 0$ and $E = 0$ that we made in Sec.\ \ref{section:Growth_rate}, therefore, the perturbations $\Phi$, $\Psi$, $\Pi^0$, $\Pi_L$, $\delta\sigma$, and $\delta$ coincide with their gauge-invariant counterparts.


\section{Coefficients in Eqs.\ (\ref{eq:eta}) and (\ref{eq:mu})}
\label{app:full_expressions}

\renewcommand{\theequation}{B\arabic{equation}}
\setcounter{equation}{0}

In this appendix, we show explicit expressions for the coefficients \{$c_{\eta 1}, c_{\eta 2}, c_{\eta\mu}, c_{\mu 1}, c_{\mu 2}$\} that appeared in Eqs.\ (\ref{eq:eta}) and (\ref{eq:mu}) for $\eta$ and $\mu$. Details of the calculation can be found in our {\tt Mathematica} supplement. The five coefficients are given by

\begin{widetext}
\bea
    c_{\eta 1} &=& 9 (r-1)^2 (X-1)^4 \left(r X^2-2 X+1\right) J^5-3 (r-1) (X-1)^2 \left(3 (r-1) X (X-1)^4+Q \left(4 X^2 (X+1)^2 r^3 \right.\right.\nn\\
    && \left.\left.-X^2 \left(7 X^2+10 X+31\right) r^2+\left(X^4+12 X^3+7 X^2+42 X-14\right) r-2 \left(X^3-4 X^2+17 X-6\right)\right)\right) J^4 \nn\\
    && -Q (X-1) \left(3 (r-1) (X-1) X \left(-X^4+8 r^3 (X+1) X^3-12 r^2 (X+3) X^3+4 X^3-42 X^2+32 X \right.\right.\nn\\
    && \left.+r \left(7 X^4+16 X^3+54 X^2-40 X+11\right)-9\right)+Q \left(4 X^2 \left(7 X^3+7 X^2-13 X-13\right) r^4 \right.\nn\\
    && -4 X^2 \left(9 X^3+27 X^2-8 X-76\right) r^3+\left(7 X^5+99 X^4+52 X^3-292 X^2-231 X+77\right) r^2 \nn\\
    && \left.\left.+\left(-14 X^4-39 X^3+9 X^2+367 X-131\right) r-3 X^3+41 X^2-141 X+55\right)\right) J^3 \hspace{4.5cm} \nn
\eea
\bea
    \hspace{1cm} && +Q \left(-12 (r-1)^2 (X-1)^2 \left((r-2) r X^3+3 X^2-3 X+1\right) X^3-Q (X-1) \left(8 X^3 \left(7 X^2-10\right) r^4 \right.\right.\nn\\
    && -4 X^3 \left(27 X^2+34 X-85\right) r^3+4 \left(15 X^5+52 X^4-64 X^3-75 X^2+47 X-11\right) r^2 \nn\\
    && \left.-\left(7 X^5+77 X^4+42 X^3-446 X^2+295 X-71\right) r+4 \left(12 X^3-39 X^2+28 X-7\right)\right) X \nn\\
    && -Q^2 \left(4 X^2 \left(4 X^4-22 X^2+21\right) r^4-4 X^2 \left(X^4+12 X^3-37 X^2-54 X+90\right) r^3\right. \nn\\
    && +4 \left(2 X^5-8 X^4-70 X^3+61 X^2+50 X-17\right) r^2+\left(7 X^4+84 X^3+30 X^2-276 X+107\right) r-48 X^2 +104 X \nn\\
    && \left.\left.-44\right)\right) J^2+4 Q^2 \left(-\left((r-1) (X-1)^2 \left(7 r^3 X^3-15 r^2 X^3-6 X^2+9 X+r \left(7 X^3+9 X^2-12 X+5\right)-4\right) X^3\right) \right.\nn\\
    && -Q (X-1) \left(\left(8 X^5-22 X^3\right) r^4-4 X^3 \left(2 X^2+4 X-17\right) r^3+\left(X^5+11 X^4-50 X^3-27 X^2+23 X-6\right) r^2 \right.\nn\\
    && \left.+\left(14 X^3+29 X^2-32 X+9\right) r-2 \left(6 X^2-7 X+2\right)\right) X+Q^2 \left(X^2 \left(8 X^2-15\right) r^4-2 X^2 \left(4 X^2+8 X-23\right) r^3 \right.\nn\\
    && \left.\left.+\left(X^4+12 X^3-32 X^2-12 X+7\right) r^2+\left(7 X^2+14 X-11\right) r-6 X+5\right)\right) J \nn\\
    && -4 Q^2 (r-1)^2 \left(-X^3+X^2+Q\right)^2 \left(X (X-1)^2+Q \left(4 r^2 X^2-1\right)\right),
\label{eq:c_eta1}
\eea
\bea
    c_{\eta 2} &=& 9 J^5 (r-1)^2 (X-1)^4 \left(r X^2-2 X+1\right)+3 J^4 (r-1) (X-1)^3 \left(Q \left(r^2 (X-13) X^2+r \left(X^3+9 X^2+28 X-14\right) \right.\right.\nn\\
    && \left.\left.-2 \left(X^2+11 X-6\right)\right)-3 (r-1) (X-1)^3 X\right)-J^3 Q (X-1)^2 \left(Q \left(2 r^3 X^2 \left(3 X^2+12 X-32\right) \right.\right.\nn\\
    && \left.+r^2 \left(-7 X^4-34 X^3+66 X^2+154 X-77\right)+r \left(14 X^3-11 X^2-236 X+131\right)+3 X^2+86 X-55\right) \nn\\
    && \left.+3 (r-1) (X-1) X \left(6 r^2 X^3-r \left(5 X^3+25 X^2-29 X+11\right)+X^3+19 X^2-23 X+9\right)\right) \nn\\
    && +J^2 Q (X-1) \left(Q^2 \left(-2 r^3 X^2 \left(2 X^3+2 X^2-19 X+23\right)+4 r^2 \left(2 X^4-7 X^3+X^2+33 X-17\right) \right.\right.\nn\\
    && \left.+r \left(7 X^3+7 X^2-169 X+107\right)+60 X-44\right)-Q X (X-1) \left(2 r^3 (9 X-17) X^3 \right.\nn\\
    && \left.-4 r^2 \left(6 X^4-4 X^3-39 X^2+36 X-11\right)+r \left(7 X^4+14 X^3-222 X^2+224 X-71\right)+72 X^2-84 X+28\right) \nn\\
    && \left.+12 (r-1)^2 X^3 (X-1)^4\right)+4 J Q^2 \left(Q^2 \left(r^3 X^2 \left(2 X^2-4 X+3\right)-r^2 \left(X^4-3 X^2+12 X-7\right) \right.\right.\nn\\
    && \left.+r (14 X-11)-6 X+5\right)-Q (X-1) X \left(2 r^3 \left(X^2-2 X+2\right) X^3-r^2 \left(X^5+X^4-6 X^3+27 X^2-23 X+6\right) \right.\nn\\
    && \left.+r \left(29 X^2-32 X+9\right)-2 \left(6 X^2-7 X+2\right)\right)+(r-1) (X-1)^2 X^3 \left(r^2 X^3+r \left(-9 X^2+12 X-5\right) \right.\nn\\
    && \left.\left.+6 X^2-9 X+4\right)\right)+4 Q^2 (r-1)^2 \left(Q-(X-1)^2 X\right) \left(Q-X^3+X^2\right)^2,
\label{eq:c_eta2}
\eea 
\bea 
    c_{\eta\mu} &=& a^2 X \left(9 J^4 (r-1)^2 (2 r+1) (X-1)^4-3 J^3 Q (r-1) (X-1)^3 \left(24 r^2-r (7 X+17)+X-1\right)\right. \nn\\
    && -J^2 Q (X-1)^2 \left(Q \left(2 r^3 \left(13 X^2+10 X-57\right)-12 r^2 \left(3 X^2+4 X-17\right)+r \left(7 X^2+34 X-77\right)-16\right)\right. \nn\\
    && \left.+36 (r-1)^2 r (X-1) X^2\right)-4 J Q^2 (X-1) \left(Q \left(r^3 \left(4 X^3-4 X^2-13 X+21\right)-r^2 \left(X^3+5 X^2-28 X+34\right)\right.\right. \nn\\
    && \left.\left.-6 r (X-1)+4\right)+(r-1) (X-1) X^2 \left(r^2 (9 X-17)+r (10-6 X)+4\right)\right) \nn\\
    && \left.+4 Q^2 (r-1) \left((X-1) X^2-Q\right) \left(Q \left(r^2 \left(-4 X^2+8 X-6\right)+r+1\right)+\left(2 r^2-r-1\right) (X-1) X^2\right)\right), 
\label{eq:c_etamu}
\eea
\bea
    c_{\mu 1} &=& 9 J^5 (r-1)^2 (X-1)^4 \left(r X^2-2 X+1\right)+3 J^4 (r-1) (X-1)^3 \left(Q \left(r^2 (X-13) X^2+r \left(X^3+9 X^2+28 X-14\right) \right.\right.\nn\\
    && \left.\left.-2 \left(X^2+11 X-6\right)\right)-3 (r-1) (X-1)^3 X\right)-J^3 Q (X-1)^2 \left(Q \left(2 r^3 X^2 \left(3 X^2+12 X-32\right) \right.\right.\nn\\
    && \left.+r^2 \left(-7 X^4-34 X^3+66 X^2+154 X-77\right)+r \left(14 X^3-11 X^2-236 X+131\right)+3 X^2+86 X-55\right) \nn\\
    && \left.+3 (r-1) (X-1) X \left(6 r^2 X^3-r \left(5 X^3+25 X^2-29 X+11\right)+X^3+19 X^2-23 X+9\right)\right) \nn\\
    && +J^2 Q (X-1) \left(Q^2 \left(-2 r^3 X^2 \left(2 X^3+2 X^2-19 X+23\right)+4 r^2 \left(2 X^4-7 X^3+X^2+33 X-17\right) \right.\right.\nn\\
    && \left.+r \left(7 X^3+7 X^2-169 X+107\right)+60 X-44\right)-Q X (X-1) \left(2 r^3 (9 X-17) X^3 \right.\nn\\
    && \left.-4 r^2 \left(6 X^4-4 X^3-39 X^2+36 X-11\right)+r \left(7 X^4+14 X^3-222 X^2+224 X-71\right)+72 X^2-84 X+28\right) \nn\\
    && \left.+12 (r-1)^2 X^3 (X-1)^4\right)+4 J Q^2 \left(Q^2 \left(r^3 X^2 \left(2 X^2-4 X+3\right)-r^2 \left(X^4-3 X^2+12 X-7\right)+r (14 X-11) \right.\right.\nn\\
    && \left.-6 X+5\right)-Q (X-1) X \left(2 r^3 \left(X^2-2 X+2\right) X^3-r^2 \left(X^5+X^4-6 X^3+27 X^2-23 X+6\right) \right.\nn\\
    && \left.+r \left(29 X^2-32 X+9\right)-2 \left(6 X^2-7 X+2\right)\right)+(r-1) (X-1)^2 X^3 \left(r^2 X^3+r \left(-9 X^2+12 X-5\right) \right.\nn\\
    && \left.\left.+6 X^2-9 X +4\right)\right)+4 Q^2 (r-1)^2 \left(Q-(X-1)^2 X\right) \left(Q-X^3+X^2\right)^2,
\label{eq:c_mu1}
\eea
\bea 
    c_{\mu 2} &=& 9 J^5 (r-1)^2 (X-1)^4 \left(r X^2-2 X+1\right)-3 J^4 (r-1) (X-1)^3 \left(Q \left(r^2 (X+15) X^2-2 r \left(2 X^3+7 X^2+14 X-7\right) \right.\right.\nn\\
    && \left.\left.+X^3+5 X^2+22 X-12\right)+3 (r-1) X (X-1)^3\right)-J^3 Q (X-1)^2 \left(Q \left(4 r^3 X^2 \left(5 X^2+6 X-21\right) \right.\right.\nn\\
    && \left.+r^2 \left(-28 X^4-48 X^3+119 X^2+154 X-77\right)+r \left(7 X^4+28 X^3-50 X^2-236 X+131\right)+9 X^2+86 X-55\right) \nn\\
    && \left.+3 (r-1) (X-1) X \left(8 r^2 X^3-r \left(9 X^3+25 X^2-29 X+11\right)+3 X^3+19 X^2-23 X+9\right)\right) \nn\\
    && +J^2 Q (X-1) \left(Q^2 \left(4 r^3 X^2 \left(-3 X^3+X^2+15 X-17\right)+4 r^2 \left(X^5+3 X^4-19 X^3+11 X^2+33 X-17\right) \right.\right.\nn\\
    && \left.+r \left(21 X^3-7 X^2-169 X+107\right)+60 X-44\right)-Q X (X-1) \left(16 r^3 (2 X-3) X^3+r^2 \left(-52 X^4+44 X^3+156 X^2 \right.\right.\nn\\
    && \left.\left.\left.-144 X+44\right)+r \left(21 X^4-222 X^2+224 X-71\right)+72 X^2-84 X+28\right)+12 (r-1)^2 X^3 (X-1)^4\right) \nn\\
    && +4 J Q^2 \left(Q^2 \left(r^3 X^2 \left(4 X^2-8 X+5\right)+r^2 \left(-3 X^4+4 X^3+X^2-12 X+7\right)+r (14 X-11)-6 X+5\right) \right.\nn\\
    && -Q (X-1) X \left(2 r^3 \left(2 X^2-4 X+3\right) X^3+r^2 \left(-3 X^5+3 X^4+4 X^3-27 X^2+23 X-6\right)+r \left(29 X^2-32 X+9\right) \right.\nn\\
    && \left.\left.-2 \left(6 X^2-7 X+2\right)\right)+(r-1) (X-1)^2 X^3 \left(r^2 X^3+r \left(-9 X^2+12 X-5\right)+6 X^2-9 X+4\right)\right) \nn\\
    && +4 Q^2 (r-1)^2 \left(Q-(X-1)^2 X\right) \left(Q-X^3+X^2\right)^2. \label{eq:c_mu2}
\eea 
\end{widetext}


\bibliography{eqmg}

\providecommand{\noopsort}[1]{}\providecommand{\singleletter}[1]{#1}%
\begin{thebibliography}{41}
\expandafter\ifx\csname natexlab\endcsname\relax\def\natexlab#1{#1}\fi
\expandafter\ifx\csname bibnamefont\endcsname\relax
  \def\bibnamefont#1{#1}\fi
\expandafter\ifx\csname bibfnamefont\endcsname\relax
  \def\bibfnamefont#1{#1}\fi
\expandafter\ifx\csname citenamefont\endcsname\relax
  \def\citenamefont#1{#1}\fi
\expandafter\ifx\csname url\endcsname\relax
  \def\url#1{\texttt{#1}}\fi
\expandafter\ifx\csname urlprefix\endcsname\relax\def\urlprefix{URL }\fi
\providecommand{\bibinfo}[2]{#2}
\providecommand{\eprint}[2][]{\url{#2}}

\bibitem[{\citenamefont{Fierz and Pauli}(1939)}]{Fierz:1939ix}
\bibinfo{author}{\bibfnamefont{M.}~\bibnamefont{Fierz}} \bibnamefont{and}
  \bibinfo{author}{\bibfnamefont{W.}~\bibnamefont{Pauli}},
  \bibinfo{journal}{Proc. Roy. Soc. Lond. A} \textbf{\bibinfo{volume}{173}},
  \bibinfo{pages}{211} (\bibinfo{year}{1939}).

\bibitem[{\citenamefont{de~Rham and Gabadadze}(2010)}]{deRham:2010ik}
\bibinfo{author}{\bibfnamefont{C.}~\bibnamefont{de~Rham}} \bibnamefont{and}
  \bibinfo{author}{\bibfnamefont{G.}~\bibnamefont{Gabadadze}},
  \bibinfo{journal}{Phys. Rev. D} \textbf{\bibinfo{volume}{82}},
  \bibinfo{pages}{044020} (\bibinfo{year}{2010}), \eprint{1007.0443}.

\bibitem[{\citenamefont{de~Rham et~al.}(2011)\citenamefont{de~Rham, Gabadadze,
  and Tolley}}]{deRham:2010kj}
\bibinfo{author}{\bibfnamefont{C.}~\bibnamefont{de~Rham}},
  \bibinfo{author}{\bibfnamefont{G.}~\bibnamefont{Gabadadze}},
  \bibnamefont{and} \bibinfo{author}{\bibfnamefont{A.~J.}
  \bibnamefont{Tolley}}, \bibinfo{journal}{Phys. Rev. Lett.}
  \textbf{\bibinfo{volume}{106}}, \bibinfo{pages}{231101}
  (\bibinfo{year}{2011}), \eprint{1011.1232}.

\bibitem[{\citenamefont{de~Rham}(2014)}]{deRham:2014zqa}
\bibinfo{author}{\bibfnamefont{C.}~\bibnamefont{de~Rham}},
  \bibinfo{journal}{Living Rev. Rel.} \textbf{\bibinfo{volume}{17}},
  \bibinfo{pages}{7} (\bibinfo{year}{2014}), \eprint{1401.4173}.

\bibitem[{\citenamefont{Hinterbichler}(2012)}]{Hinterbichler:2011tt}
\bibinfo{author}{\bibfnamefont{K.}~\bibnamefont{Hinterbichler}},
  \bibinfo{journal}{Rev. Mod. Phys.} \textbf{\bibinfo{volume}{84}},
  \bibinfo{pages}{671} (\bibinfo{year}{2012}), \eprint{1105.3735}.

\bibitem[{\citenamefont{Boulware and Deser}(1972)}]{Boulware:1972yco}
\bibinfo{author}{\bibfnamefont{D.~G.} \bibnamefont{Boulware}} \bibnamefont{and}
  \bibinfo{author}{\bibfnamefont{S.}~\bibnamefont{Deser}},
  \bibinfo{journal}{Phys. Rev. D} \textbf{\bibinfo{volume}{6}},
  \bibinfo{pages}{3368} (\bibinfo{year}{1972}).

\bibitem[{\citenamefont{D'Amico et~al.}(2011)\citenamefont{D'Amico, de~Rham,
  Dubovsky, Gabadadze, Pirtskhalava, and Tolley}}]{DAmico:2011eto}
\bibinfo{author}{\bibfnamefont{G.}~\bibnamefont{D'Amico}},
  \bibinfo{author}{\bibfnamefont{C.}~\bibnamefont{de~Rham}},
  \bibinfo{author}{\bibfnamefont{S.}~\bibnamefont{Dubovsky}},
  \bibinfo{author}{\bibfnamefont{G.}~\bibnamefont{Gabadadze}},
  \bibinfo{author}{\bibfnamefont{D.}~\bibnamefont{Pirtskhalava}},
  \bibnamefont{and} \bibinfo{author}{\bibfnamefont{A.~J.}
  \bibnamefont{Tolley}}, \bibinfo{journal}{Phys. Rev. D}
  \textbf{\bibinfo{volume}{84}}, \bibinfo{pages}{124046}
  (\bibinfo{year}{2011}), \eprint{1108.5231}.

\bibitem[{\citenamefont{D'Amico et~al.}(2013)\citenamefont{D'Amico, Gabadadze,
  Hui, and Pirtskhalava}}]{DAmico:2012hia}
\bibinfo{author}{\bibfnamefont{G.}~\bibnamefont{D'Amico}},
  \bibinfo{author}{\bibfnamefont{G.}~\bibnamefont{Gabadadze}},
  \bibinfo{author}{\bibfnamefont{L.}~\bibnamefont{Hui}}, \bibnamefont{and}
  \bibinfo{author}{\bibfnamefont{D.}~\bibnamefont{Pirtskhalava}},
  \bibinfo{journal}{Phys. Rev. D} \textbf{\bibinfo{volume}{87}},
  \bibinfo{pages}{064037} (\bibinfo{year}{2013}), \eprint{1206.4253}.

\bibitem[{\citenamefont{Gannouji et~al.}(2013)\citenamefont{Gannouji, Hossain,
  Sami, and Saridakis}}]{Gannouji:2013rwa}
\bibinfo{author}{\bibfnamefont{R.}~\bibnamefont{Gannouji}},
  \bibinfo{author}{\bibfnamefont{M.~W.} \bibnamefont{Hossain}},
  \bibinfo{author}{\bibfnamefont{M.}~\bibnamefont{Sami}}, \bibnamefont{and}
  \bibinfo{author}{\bibfnamefont{E.~N.} \bibnamefont{Saridakis}},
  \bibinfo{journal}{Phys. Rev. D} \textbf{\bibinfo{volume}{87}},
  \bibinfo{pages}{123536} (\bibinfo{year}{2013}), \eprint{1304.5095}.

\bibitem[{\citenamefont{G\"umr\"uk\c{c}\"uo\u{g}lu
  et~al.}(2013)\citenamefont{G\"umr\"uk\c{c}\"uo\u{g}lu, Hinterbichler, Lin,
  Mukohyama, and Trodden}}]{Gumrukcuoglu:2013nza}
\bibinfo{author}{\bibfnamefont{A.~E.}
  \bibnamefont{G\"umr\"uk\c{c}\"uo\u{g}lu}},
  \bibinfo{author}{\bibfnamefont{K.}~\bibnamefont{Hinterbichler}},
  \bibinfo{author}{\bibfnamefont{C.}~\bibnamefont{Lin}},
  \bibinfo{author}{\bibfnamefont{S.}~\bibnamefont{Mukohyama}},
  \bibnamefont{and} \bibinfo{author}{\bibfnamefont{M.}~\bibnamefont{Trodden}},
  \bibinfo{journal}{Phys. Rev. D} \textbf{\bibinfo{volume}{88}},
  \bibinfo{pages}{024023} (\bibinfo{year}{2013}), \eprint{1304.0449}.

\bibitem[{\citenamefont{De~Felice and Mukohyama}(2014)}]{DeFelice:2013tsa}
\bibinfo{author}{\bibfnamefont{A.}~\bibnamefont{De~Felice}} \bibnamefont{and}
  \bibinfo{author}{\bibfnamefont{S.}~\bibnamefont{Mukohyama}},
  \bibinfo{journal}{Phys. Lett. B} \textbf{\bibinfo{volume}{728}},
  \bibinfo{pages}{622} (\bibinfo{year}{2014}), \eprint{1306.5502}.

\bibitem[{\citenamefont{Kahniashvili et~al.}(2015)\citenamefont{Kahniashvili,
  Kar, Lavrelashvili, Agarwal, Heisenberg, and
  Kosowsky}}]{Kahniashvili:2014wua}
\bibinfo{author}{\bibfnamefont{T.}~\bibnamefont{Kahniashvili}},
  \bibinfo{author}{\bibfnamefont{A.}~\bibnamefont{Kar}},
  \bibinfo{author}{\bibfnamefont{G.}~\bibnamefont{Lavrelashvili}},
  \bibinfo{author}{\bibfnamefont{N.}~\bibnamefont{Agarwal}},
  \bibinfo{author}{\bibfnamefont{L.}~\bibnamefont{Heisenberg}},
  \bibnamefont{and} \bibinfo{author}{\bibfnamefont{A.}~\bibnamefont{Kosowsky}},
  \bibinfo{journal}{Phys. Rev. D} \textbf{\bibinfo{volume}{91}},
  \bibinfo{pages}{041301} (\bibinfo{year}{2015}), \bibinfo{note}{[Erratum:
  Phys.Rev.D 100, 089902 (2019)]}, \eprint{1412.4300}.

\bibitem[{\citenamefont{Motohashi and Hu}(2014)}]{Motohashi:2014una}
\bibinfo{author}{\bibfnamefont{H.}~\bibnamefont{Motohashi}} \bibnamefont{and}
  \bibinfo{author}{\bibfnamefont{W.}~\bibnamefont{Hu}}, \bibinfo{journal}{Phys.
  Rev. D} \textbf{\bibinfo{volume}{90}}, \bibinfo{pages}{104008}
  (\bibinfo{year}{2014}), \eprint{1408.4813}.

\bibitem[{\citenamefont{Heisenberg}(2015)}]{Heisenberg:2015voa}
\bibinfo{author}{\bibfnamefont{L.}~\bibnamefont{Heisenberg}},
  \bibinfo{journal}{JCAP} \textbf{\bibinfo{volume}{04}}, \bibinfo{pages}{010}
  (\bibinfo{year}{2015}), \eprint{1501.07796}.

\bibitem[{\citenamefont{Gümrükçüoğlu
  et~al.}(2016)\citenamefont{Gümrükçüoğlu, Koyama, and
  Mukohyama}}]{Gumrukcuoglu:2016hic}
\bibinfo{author}{\bibfnamefont{A.~E.} \bibnamefont{Gümrükçüoğlu}},
  \bibinfo{author}{\bibfnamefont{K.}~\bibnamefont{Koyama}}, \bibnamefont{and}
  \bibinfo{author}{\bibfnamefont{S.}~\bibnamefont{Mukohyama}},
  \bibinfo{journal}{Phys. Rev. D} \textbf{\bibinfo{volume}{94}},
  \bibinfo{pages}{123510} (\bibinfo{year}{2016}), \eprint{1610.03562}.

\bibitem[{\citenamefont{Kluso\v{n}}(2014)}]{Kluson:2013jea}
\bibinfo{author}{\bibfnamefont{J.}~\bibnamefont{Kluso\v{n}}},
  \bibinfo{journal}{J. Grav.} \textbf{\bibinfo{volume}{2014}},
  \bibinfo{pages}{413835} (\bibinfo{year}{2014}), \eprint{1309.0956}.

\bibitem[{\citenamefont{Mukohyama}(2017)}]{Mukohyama:2013raa}
\bibinfo{author}{\bibfnamefont{S.}~\bibnamefont{Mukohyama}},
  \bibinfo{journal}{Phys. Rev. D} \textbf{\bibinfo{volume}{96}},
  \bibinfo{pages}{044029} (\bibinfo{year}{2017}), \eprint{1309.2146}.

\bibitem[{\citenamefont{Anselmi et~al.}(2017)\citenamefont{Anselmi, Kumar,
  L\'opez~Nacir, and Starkman}}]{Anselmi:2017hwr}
\bibinfo{author}{\bibfnamefont{S.}~\bibnamefont{Anselmi}},
  \bibinfo{author}{\bibfnamefont{S.}~\bibnamefont{Kumar}},
  \bibinfo{author}{\bibfnamefont{D.}~\bibnamefont{L\'opez~Nacir}},
  \bibnamefont{and} \bibinfo{author}{\bibfnamefont{G.~D.}
  \bibnamefont{Starkman}}, \bibinfo{journal}{Phys. Rev. D}
  \textbf{\bibinfo{volume}{96}}, \bibinfo{pages}{084001}
  (\bibinfo{year}{2017}), \eprint{1706.01872}.

\bibitem[{\citenamefont{Golovnev and Trukhin}(2017)}]{Golovnev:2017zxk}
\bibinfo{author}{\bibfnamefont{A.}~\bibnamefont{Golovnev}} \bibnamefont{and}
  \bibinfo{author}{\bibfnamefont{A.}~\bibnamefont{Trukhin}},
  \bibinfo{journal}{Phys. Rev. D} \textbf{\bibinfo{volume}{96}},
  \bibinfo{pages}{104032} (\bibinfo{year}{2017}), \eprint{1706.07215}.

\bibitem[{\citenamefont{Gümrükçüoğlu
  et~al.}(2017)\citenamefont{Gümrükçüoğlu, Koyama, and
  Mukohyama}}]{Gumrukcuoglu:2017ioy}
\bibinfo{author}{\bibfnamefont{A.~E.} \bibnamefont{Gümrükçüoğlu}},
  \bibinfo{author}{\bibfnamefont{K.}~\bibnamefont{Koyama}}, \bibnamefont{and}
  \bibinfo{author}{\bibfnamefont{S.}~\bibnamefont{Mukohyama}},
  \bibinfo{journal}{Phys. Rev. D} \textbf{\bibinfo{volume}{96}},
  \bibinfo{pages}{044041} (\bibinfo{year}{2017}), \eprint{1707.02004}.

\bibitem[{\citenamefont{Solomon et~al.}(2014)\citenamefont{Solomon, Akrami, and
  Koivisto}}]{Solomon:2014dua}
\bibinfo{author}{\bibfnamefont{A.~R.} \bibnamefont{Solomon}},
  \bibinfo{author}{\bibfnamefont{Y.}~\bibnamefont{Akrami}}, \bibnamefont{and}
  \bibinfo{author}{\bibfnamefont{T.~S.} \bibnamefont{Koivisto}},
  \bibinfo{journal}{JCAP} \textbf{\bibinfo{volume}{10}}, \bibinfo{pages}{066}
  (\bibinfo{year}{2014}), \eprint{1404.4061}.

\bibitem[{\citenamefont{Kenna-Allison et~al.}(2020)\citenamefont{Kenna-Allison,
  Gümrükçüoğlu, and Koyama}}]{Kenna-Allison:2020egn}
\bibinfo{author}{\bibfnamefont{M.}~\bibnamefont{Kenna-Allison}},
  \bibinfo{author}{\bibfnamefont{A.~E.} \bibnamefont{Gümrükçüoğlu}},
  \bibnamefont{and} \bibinfo{author}{\bibfnamefont{K.}~\bibnamefont{Koyama}},
  \bibinfo{journal}{Phys. Rev. D} \textbf{\bibinfo{volume}{102}},
  \bibinfo{pages}{103524} (\bibinfo{year}{2020}), \eprint{2009.05405}.

\bibitem[{\citenamefont{Manita and Kimura}(2022)}]{Manita:2022}
\bibinfo{author}{\bibfnamefont{Y.}~\bibnamefont{Manita}} \bibnamefont{and}
  \bibinfo{author}{\bibfnamefont{R.}~\bibnamefont{Kimura}},
  \bibinfo{journal}{Phys. Rev. D} \textbf{\bibinfo{volume}{105}},
  \bibinfo{pages}{084038} (\bibinfo{year}{2022}),
  \urlprefix\url{https://link.aps.org/doi/10.1103/PhysRevD.105.084038}.

\bibitem[{\citenamefont{De~Felice
  et~al.}(2021{\natexlab{a}})\citenamefont{De~Felice, Mukohyama, and
  Pookkillath}}]{DeFelice:2021trp}
\bibinfo{author}{\bibfnamefont{A.}~\bibnamefont{De~Felice}},
  \bibinfo{author}{\bibfnamefont{S.}~\bibnamefont{Mukohyama}},
  \bibnamefont{and} \bibinfo{author}{\bibfnamefont{M.~C.}
  \bibnamefont{Pookkillath}}, \bibinfo{journal}{JCAP}
  \textbf{\bibinfo{volume}{12}}, \bibinfo{pages}{011}
  (\bibinfo{year}{2021}{\natexlab{a}}), \eprint{2110.01237}.

\bibitem[{\citenamefont{Zhang}(2006)}]{Zhang:2005vt}
\bibinfo{author}{\bibfnamefont{P.}~\bibnamefont{Zhang}},
  \bibinfo{journal}{Phys. Rev. D} \textbf{\bibinfo{volume}{73}},
  \bibinfo{pages}{123504} (\bibinfo{year}{2006}), \eprint{astro-ph/0511218}.

\bibitem[{\citenamefont{Bean et~al.}(2007)\citenamefont{Bean, Bernat, Pogosian,
  Silvestri, and Trodden}}]{Bean:2006up}
\bibinfo{author}{\bibfnamefont{R.}~\bibnamefont{Bean}},
  \bibinfo{author}{\bibfnamefont{D.}~\bibnamefont{Bernat}},
  \bibinfo{author}{\bibfnamefont{L.}~\bibnamefont{Pogosian}},
  \bibinfo{author}{\bibfnamefont{A.}~\bibnamefont{Silvestri}},
  \bibnamefont{and} \bibinfo{author}{\bibfnamefont{M.}~\bibnamefont{Trodden}},
  \bibinfo{journal}{Phys. Rev. D} \textbf{\bibinfo{volume}{75}},
  \bibinfo{pages}{064020} (\bibinfo{year}{2007}), \eprint{astro-ph/0611321}.

\bibitem[{\citenamefont{Higuchi}(1987)}]{Higuchi:1986py}
\bibinfo{author}{\bibfnamefont{A.}~\bibnamefont{Higuchi}},
  \bibinfo{journal}{Nucl. Phys. B} \textbf{\bibinfo{volume}{282}},
  \bibinfo{pages}{397} (\bibinfo{year}{1987}).

\bibitem[{\citenamefont{Fasiello and Tolley}(2012)}]{Fasiello_2012}
\bibinfo{author}{\bibfnamefont{M.}~\bibnamefont{Fasiello}} \bibnamefont{and}
  \bibinfo{author}{\bibfnamefont{A.~J.} \bibnamefont{Tolley}},
  \bibinfo{journal}{JCAP} \textbf{\bibinfo{volume}{11}}, \bibinfo{pages}{035}
  (\bibinfo{year}{2012}), \eprint{1206.3852}.

\bibitem[{\citenamefont{Hojjati et~al.}(2012)\citenamefont{Hojjati, Pogosian,
  Silvestri, and Talbot}}]{Hojjati:2012rf}
\bibinfo{author}{\bibfnamefont{A.}~\bibnamefont{Hojjati}},
  \bibinfo{author}{\bibfnamefont{L.}~\bibnamefont{Pogosian}},
  \bibinfo{author}{\bibfnamefont{A.}~\bibnamefont{Silvestri}},
  \bibnamefont{and} \bibinfo{author}{\bibfnamefont{S.}~\bibnamefont{Talbot}},
  \bibinfo{journal}{Phys. Rev. D} \textbf{\bibinfo{volume}{86}},
  \bibinfo{pages}{123503} (\bibinfo{year}{2012}), \eprint{1210.6880}.

\bibitem[{\citenamefont{Sawicki and Bellini}(2015)}]{Sawicki:2015zya}
\bibinfo{author}{\bibfnamefont{I.}~\bibnamefont{Sawicki}} \bibnamefont{and}
  \bibinfo{author}{\bibfnamefont{E.}~\bibnamefont{Bellini}},
  \bibinfo{journal}{Phys. Rev. D} \textbf{\bibinfo{volume}{92}},
  \bibinfo{pages}{084061} (\bibinfo{year}{2015}), \eprint{1503.06831}.

\bibitem[{\citenamefont{Baker et~al.}(2021)}]{Baker:2019gxo}
\bibinfo{author}{\bibfnamefont{T.}~\bibnamefont{Baker}} \bibnamefont{et~al.},
  \bibinfo{journal}{Rev. Mod. Phys.} \textbf{\bibinfo{volume}{93}},
  \bibinfo{pages}{015003} (\bibinfo{year}{2021}), \eprint{1908.03430}.

\bibitem[{\citenamefont{Ade et~al.}(2016)}]{Planck:2015fie}
\bibinfo{author}{\bibfnamefont{P.~A.~R.} \bibnamefont{Ade}}
  \bibnamefont{et~al.}, \bibinfo{journal}{Astron. Astrophys.}
  \textbf{\bibinfo{volume}{594}}, \bibinfo{pages}{A13} (\bibinfo{year}{2016}),
  \eprint{1502.01589}.

\bibitem[{\citenamefont{Magaña et~al.}(2012)\citenamefont{Magaña, Matos,
  Robles, and Suárez}}]{Magana:2012}
\bibinfo{author}{\bibfnamefont{J.}~\bibnamefont{Magaña}},
  \bibinfo{author}{\bibfnamefont{T.}~\bibnamefont{Matos}},
  \bibinfo{author}{\bibfnamefont{V.}~\bibnamefont{Robles}}, \bibnamefont{and}
  \bibinfo{author}{\bibfnamefont{A.}~\bibnamefont{Suárez}}
  (\bibinfo{year}{2012}), \eprint{1201.6107}.

\bibitem[{\citenamefont{Ure\~na L\'opez}(2019)}]{Urena-Lopez:2019kud}
\bibinfo{author}{\bibfnamefont{L.~A.} \bibnamefont{Ure\~na L\'opez}},
  \bibinfo{journal}{Front. Astron. Space Sci.} \textbf{\bibinfo{volume}{6}},
  \bibinfo{pages}{47} (\bibinfo{year}{2019}).

\bibitem[{\citenamefont{Turner}(1983)}]{Turner:1983he}
\bibinfo{author}{\bibfnamefont{M.~S.} \bibnamefont{Turner}},
  \bibinfo{journal}{Phys. Rev. D} \textbf{\bibinfo{volume}{28}},
  \bibinfo{pages}{1243} (\bibinfo{year}{1983}).

\bibitem[{\citenamefont{Suzuki et~al.}(2012)}]{Suzuki:2011hu}
\bibinfo{author}{\bibfnamefont{N.}~\bibnamefont{Suzuki}} \bibnamefont{et~al.},
  \bibinfo{journal}{Astrophys. J.} \textbf{\bibinfo{volume}{746}},
  \bibinfo{pages}{85} (\bibinfo{year}{2012}), \eprint{1105.3470}.

\bibitem[{\citenamefont{Perenon et~al.}(2019)\citenamefont{Perenon, Bel,
  Maartens, and de~la Cruz-Dombriz}}]{Perenon:2019dpc}
\bibinfo{author}{\bibfnamefont{L.}~\bibnamefont{Perenon}},
  \bibinfo{author}{\bibfnamefont{J.}~\bibnamefont{Bel}},
  \bibinfo{author}{\bibfnamefont{R.}~\bibnamefont{Maartens}}, \bibnamefont{and}
  \bibinfo{author}{\bibfnamefont{A.}~\bibnamefont{de~la Cruz-Dombriz}},
  \bibinfo{journal}{JCAP} \textbf{\bibinfo{volume}{06}}, \bibinfo{pages}{020}
  (\bibinfo{year}{2019}), \eprint{1901.11063}.

\bibitem[{\citenamefont{G\"umr\"uk\c{c}\"uoglu and
  Koyama}(2019)}]{Gumrukcuoglu:2019rsw}
\bibinfo{author}{\bibfnamefont{A.~E.} \bibnamefont{G\"umr\"uk\c{c}\"uoglu}}
  \bibnamefont{and} \bibinfo{author}{\bibfnamefont{K.}~\bibnamefont{Koyama}},
  \bibinfo{journal}{Phys. Rev. D} \textbf{\bibinfo{volume}{99}},
  \bibinfo{pages}{084004} (\bibinfo{year}{2019}), \eprint{1902.01391}.

\bibitem[{\citenamefont{De~Felice
  et~al.}(2021{\natexlab{b}})\citenamefont{De~Felice, Larrouturou, Mukohyama,
  and Oliosi}}]{DeFelice:2020ecp}
\bibinfo{author}{\bibfnamefont{A.}~\bibnamefont{De~Felice}},
  \bibinfo{author}{\bibfnamefont{F.}~\bibnamefont{Larrouturou}},
  \bibinfo{author}{\bibfnamefont{S.}~\bibnamefont{Mukohyama}},
  \bibnamefont{and} \bibinfo{author}{\bibfnamefont{M.}~\bibnamefont{Oliosi}},
  \bibinfo{journal}{JCAP} \textbf{\bibinfo{volume}{04}}, \bibinfo{pages}{015}
  (\bibinfo{year}{2021}{\natexlab{b}}), \eprint{2012.01073}.

\bibitem[{\citenamefont{Gümrükçüoğlu
  et~al.}(2020)\citenamefont{Gümrükçüoğlu, Kimura, and
  Koyama}}]{Gumrukcuoglu:2020utx}
\bibinfo{author}{\bibfnamefont{A.~E.} \bibnamefont{Gümrükçüoğlu}},
  \bibinfo{author}{\bibfnamefont{R.}~\bibnamefont{Kimura}}, \bibnamefont{and}
  \bibinfo{author}{\bibfnamefont{K.}~\bibnamefont{Koyama}},
  \bibinfo{journal}{Phys. Rev. D} \textbf{\bibinfo{volume}{101}},
  \bibinfo{pages}{124021} (\bibinfo{year}{2020}), \eprint{2003.11831}.

\bibitem[{\citenamefont{Akbarieh et~al.}(2021)\citenamefont{Akbarieh,
  Kazempour, and Shao}}]{Akbarieh:2021vhv}
\bibinfo{author}{\bibfnamefont{A.~R.} \bibnamefont{Akbarieh}},
  \bibinfo{author}{\bibfnamefont{S.}~\bibnamefont{Kazempour}},
  \bibnamefont{and} \bibinfo{author}{\bibfnamefont{L.}~\bibnamefont{Shao}},
  \bibinfo{journal}{Phys. Rev. D} \textbf{\bibinfo{volume}{103}},
  \bibinfo{pages}{123518} (\bibinfo{year}{2021}), \eprint{2105.03744}.

\end{thebibliography}

\end{document}